\documentclass[a4paper,11pt,nontitlepage]{article}
\usepackage{jheppub} 
\pdfoutput=1

\RequirePackage[numbers,sort&compress]{natbib}

\RequirePackage[utf8]{inputenc} 			
\RequirePackage[T1]{fontenc} 				
\RequirePackage[autostyle=true]{csquotes}	
\RequirePackage{graphicx}					
\RequirePackage{listings}					
\RequirePackage{xcolor}					    
\RequirePackage{siunitx}   					
\RequirePackage[english]{babel} 
\RequirePackage{hyperref}					
\RequirePackage[all]{hypcap}				
\RequirePackage{amsmath}
\RequirePackage{physics}					
\RequirePackage[varg]{txfonts}       		
\RequirePackage{caption}					
\RequirePackage{subcaption}
\usepackage{mathtools}
\RequirePackage{pgf}

\RequirePackage{booktabs, multirow}

\graphicspath{figs/}

\newcommand{\nuclide}[2]{{}^{#1}{\textnormal{#2}}}

\newcommand{\triton}{\ensuremath{{ }^{3}\mathrm{H}}}
\newcommand{\He}[1]{\ensuremath{{ }^{#1}\mathrm{He}}}
\newcommand{\Li}[1]{\ensuremath{{ }^{#1}\mathrm{Li}}}
\newcommand{\Be}[1]{\ensuremath{{ }^{#1}\mathrm{Be}}}

\title{Bounds on variations of the strange quark mass  from  Big Bang nucleosynthesis}

\author[a,b,c]{Ulf-G. Mei{\ss}ner,}
\author[b,a]{Bernard Metsch,}
\author[a]{and Helen Meyer}

\affiliation[a]{Helmholtz-Institut f\"{u}r Strahlen- und Kernphysik and Bethe Center for Theoretical Physics,\\
  Universit\"{a}t Bonn, D-53115 Bonn, Germany}
\affiliation[b]{Institute for Advanced Simulation (IAS-4), Forschungszentrum J\"ulich, D-52425 J\"ulich, Germany}
\affiliation[c]{Peng Huanwu Collaborative Center for Research and Education, Beihang University, Beijing 100191,
  China}

\emailAdd{meissner@hiskp.uni-bonn.de}
\emailAdd{metsch@hiskp.uni-bonn.de}
\emailAdd{hmeyer@hiskp.uni-bonn.de} 

\abstract{
We analyze the effect of a variation of the strange nucleon matrix element $\langle N| m_s \bar{s}s|N\rangle$
on the abundances of the light elements produced in the Big Bang. For that, we vary the 
nucleon mass in the leading eight reactions that involve neutrons, protons and the four
lightest nuclei. We use various available Big Bang nucleosynthesis codes and find that
the measured deuterium and $^4$He abundances set strict limits on the nucleon mass variations.
This translates into an upper bound of possible variations of the strange quark mass,
$\left| \Delta m_s/m_s \right| \leq 5.1\%$.
}

\begin{document}

\maketitle	


\section{Introduction}
Big Bang nucleosynthesis (BBN) is a fine laboratory for studying possible
variations of the fundamental constants of nature as well as to search for new physics effects, see e.g.
\cite{Hogan:1999wh,Uzan:2002vq,Schellekens:2013bpa,Meissner:2014pma,Donoghue:2016tjk,Adams:2019kby,Cyburt:2004yc,Pospelov:2010hj,Anagnostopoulos:2022gej,Burns:2023sgx,Uzan:2024ded}.
Here, we will consider the first issue, namely the bounds  on variations
of certain parameters from the primordial element abundances.
Mostly, one has studied the variation of the electromagnetic
fine-structure constant $\alpha_{\rm EM}$ (for a recent work with many references,
see~\cite{Meissner:2023voo}) or of the light quark masses
$m_u, m_d$ (or alternatively, the Higgs vacuum expectation value (VEV)
$v$), for very recent works see Refs.~\cite{Burns:2024ods,Meyer:2024auq}.

However, the nucleon and thus the nuclei contain a certain
amount of strangeness 
as indicated by the non-vanishing matrix elements $\langle N| m_s
\bar{s}s|N\rangle$ or $\langle N| m_s \bar{s} \gamma_\mu s|N\rangle$,
just to mention
two~\cite{Maas:2017snj,FlavourLatticeAveragingGroupFLAG:2024oxs}. 
Since the nuclei that take part in the BBN reaction network are made of
neutrons and protons, their masses are also sensitive to the amount
of strangeness in the nucleon and similarly, the mass differences of the pertinent reactions
depend on the precise value of the matrix-element $\langle N| m_s \bar{s}s|N\rangle$.
Such effects have not been considered before. Note that we are not talking about
hypernuclei here, which are nuclei that contain one or two hyperons.

There are many other possible sources of strangeness in the BBN reaction
network. First, kaon loops could influence the vector coupling
constant that plays a role in neutron $\beta$-decay. Such effects have
been shown to be tiny~\cite{Kaiser:2001yc} and can thus be
neglected. Further, in the boson-exchange models of the nuclear force that
underlie the modeling of atomic nuclei, strangeness could enter via
$\eta$-exchanges or $\pi K$, $KK$ loops. While the $\eta$ coupling to
the nucleon is very small and such effects can be neglected,  see e.g.~\cite{Jain:1987sz}
and references therein, loops
involving kaons are of such short distance, that they can effectively
be absorbed in the four-nucleon contact interactions that are used in
the modern theory of nuclear forces~\cite{Epelbaum:2008ga}. As it
turns out, the dominant strangeness effect in BBN is due to the mass shift
$\langle N| m_s \bar{s}s|N\rangle$, which enters the nucleon mass via
the trace anomaly of the energy-momentum tensor. As will be shown
below, this matrix element is of the order of 50~MeV, and such a mass
shift can strongly influence the BBN network.

The manuscript is organized as follows. In Sect.~\ref{sec:strange} we
discuss the determination of the matrix element $\langle N| m_s
\bar{s}s|N\rangle$ and give the bounds on its value.  Sect.~\ref{sec:calc}
contains the details of the calculational procedure. First, we consider
the nucleon mass dependence of the leading eight reactions in the BBN
network, making use of results from pionless EFT as well as Nuclear Lattice
EFT (NLEFT). Second, we discuss the modeling of the nucleon mass dependence of 
nuclear reaction rates. We also present the temperature-dependent reaction
rates based on various inputs from NLEFT and/or pionless EFT. This we
consider the main systematic error of the calculation.
In Sect.~\ref{sec:res}, we present and and discuss our results on the nucleon mass
variations consistent with the observed abundances and the consequences for the
strangeness content of the nucleon.
Some technicalities are relegated to the appendices.

\section{Strangeness in the nucleon}
\label{sec:strange}

The quark mass contribution to the nucleon mass (we neglect the heavy
flavors $c,b$ here) can be derived from the mass term of the three-flavor QCD Hamiltonian,
\begin{equation}
  \label{eq:qcdmass}
  {\cal H}_{\rm QCD}^{\rm mass}
  =
  m_u \,\bar u u + m_d  \,\bar  d d + m_s \, \bar ss~.
\end{equation}
The trace anomaly of the energy-momentum tensor $\Theta_{\mu\nu}$ allows one to
quantify the effect of the quark mass terms on the nucleon mass, see e.g.
Refs.~\cite{Collins:1976yq,Crewther:1972kn,Nielsen:1977sy},
\begin{equation}
\label{eq:trace}  
\langle N | \Theta_\mu^\mu |N\rangle =
\left\langle\, N\,\left| -\frac{\beta(g)}{g^3}\, G_{\mu\nu}^aG^{a,\mu\nu}\right|\,N\,\right\rangle  
+ \langle N| m_u\bar uu + m_d \bar dd + m_s \bar ss |N\rangle
\end{equation}
where $|N\rangle$ is the nucleon spinor,
$\beta$ is the QCD $\beta$-function, $g$ the strong coupling constant, 
$G_{\mu\nu}^a$ the non-abelian gluon field strength tensor, $a=1,...,8$
is a color index and  we have neglected the anomalous dimension contribution to the
quark mass term for simplicity. The first term related to the QCD gauge fields in Eq.~\eqref{eq:trace} generates
the bulk of the nucleon mass, that is the nucleon mass in the three-flavor chiral limit,
the famous ``mass without mass''~\cite{Wheeler:1955zz,Wilczek:1999}.
The quark mass contribution to the nucleon mass is encoded in the second
term in in Eq.~\eqref{eq:trace}, where the first two terms from the light quarks $u,d$ are related to the
well-known pion-nucleon sigma term, $\sigma_{\pi N}$, that can be precisely determined from
pion-nucleon scattering data using Roy-Steiner equations.  Its most
recent and accurate determination including isospin-breaking effects
gives
$\sigma_{\pi N} = 59.0(3.5)\,$MeV~\cite{Hoferichter:2023ptl}.
The strange sigma-term, that gives the strangeness contribution of the nucleon mass,
\begin{equation}
  \sigma_{s}
  =
  \langle N| m_s\, \bar{s}s|N\rangle~,
\end{equation}
is  known much less precisely and can be determined using lattice QCD. The FLAG
group quotes an average of
$\sigma_{s} = 44.9(6.4)\,$MeV
for simulations with $N_f = 2+1$ flavors  and 
$\sigma_{s} = 41.0(8.8)\,$MeV
for simulations with $N_f = 2+1+1$ 
flavors~\cite{FlavourLatticeAveragingGroupFLAG:2024oxs}.  However,
looking at the individual results that enter these averages, one sees that these
span a range from $-40$ to  $160$~MeV within the given uncertainties. We will
therefore vary this matrix element in the range from $-94$ to $94$~MeV,
corresponding to changes of at most 10\% of the nucleon mass.
This variation will not only modify the nucleon mass (the proton and
the neutron mass equally, since we are dealing with an isoscalar
operator) but consequently also the masses of the involved nuclei. This, in turn, affects the $Q$-values, meaning the differences between the masses of incoming and outgoing particles, of the various reactions in the network. From imposing consistency with experimental data for primordial
abundances, one can then also get bounds on the possible variations of
the strange quark mass, if one assumes that the strange condensate in
the nucleon remains unchanged. Note, however, that the quark masses
and the condensate are scale- and scheme-dependent quantities, so that
only their product is an observable. Thus, it is customary to use the
$\overline{\rm MS}$ scheme at the scale $\mu= 2\,$GeV when one
discusses the light quark masses. In what follows, we will vary the
strange sigma-term and thus the nucleon mass and derive bounds on such
variations from the observed primordial abundances. More precisely,
if we assume the strange quark condensate to be constant, we can 
extract a possible variation of the strange quark mass from the 
nucleon mass variation via
\begin{equation}
\left|\delta m_s\right| 
=
\left|\frac{\Delta m_s}{m_s}\right| 
= 
\frac{1}{\sigma_s}\, \left|\frac{\Delta m_N}{m_N}\right|
=
\frac{1}{\sigma_s}\,\left|\delta_{m_N}\right|~.
\end{equation}



\section{Calculational procedure}
\label{sec:calc}

In this section, we discuss how changes in the nucleon mass affect the BBN reaction network.
We heavily borrow from the formalism laid out in Refs.~\cite{Meissner:2022dhg,Meissner:2023voo}
and present here only the details required to investigate the influence of nucleon mass shifts
on the primordial abundances.

\subsection{Nucleon mass dependence of binding energies and other quantities}\label{ss:BE}

Fortunately, for the leading nuclear reaction $n + p \to d + \gamma$ a
detailed and accurate theoretical description within the framework of
pionless Effective Field Theory (EFT) exists, see
Refs.~\cite{Chen:1999bg,Rupak:1999rk}. Accordingly, one can calculate
the variation of the cross section with a variation of the nucleon
mass $m_N$ and the leading effective range parameters for
$np$-scattering: $a_s, r_s, a_t, r_t$ denoting the singlet ($^1S_0$) scattering
length, the corresponding effective range, the triplet ($^3S_1$) scattering
length and corresponding effective range, respectively.

In order to determine the variation of the effective range parameters
with a variation of the nucleon mass we calculated the low energy $np$
scattering on the basis on the next-to-next-to-next-to leading order (N$^3$LO) chiral effective field theory Hamiltonian implemented on the lattice consisting of the one-pion exchange
potential and various contact interactions to account for the
short range behavior, similar to what was done in earlier works~\cite{Epelbaum:2012iu,Epelbaum:2013wla,Elhatisari:2021eyg}. The calculation was done on a three-dimensional lattice with $L=45$ (in lattice units) in each spatial dimension with a lattice spacing $a = \SI{1.32}{\femto\m}$. Note that in the framework of Nuclear Lattice Effective Field Theory (NLEFT) the lattice spacing works as a regulator and is not meant to approach zero. The contact operators at given orders are listed, e.g., in \cite{Li:2018ymw} and the corresponding low-energy constants (LECs) were found from a fit to data for the phase shifts in the $^1S_0$ and ${}^3S_1-{}^3D_1$ channels. 



For comparison we performed the same calculation with a simple square-well potential, 
as earlier done e.g. in Ref.~\cite{Ubaldi:2008nf},
where the width and depth were adjusted to experimental values at the nominal 
nucleon mass. For details, see Appendix~\ref{app:square}.

Note that in both calculations we assume the interaction to be nucleon-mass 
independent and the variation of the scattering parameters is due to the 
change of the reduced mass in the kinetic energy only. Varying the strange quark matrix element between $\SI{-94}{\mega\electronvolt}$ to $\SI{94}{\mega\electronvolt}$ as described in Sect.~\ref{sec:strange}, is roughly equivalent to changing the nucleon mass by $\pm 10\%$, so we define
\begin{equation}\label{eq:deltamN}
  m_N(\delta_{m_N}) = m_N\,\cdot (1+\delta_{m_N})~,  
\end{equation}
where $m_N = \SI{938.92}{\mega\electronvolt}$ is the nominal value of the average nucleon mass~\cite{ParticleDataGroup:2024cfk}.
 
\begin{figure}[!htb]
  \centering
  \includegraphics{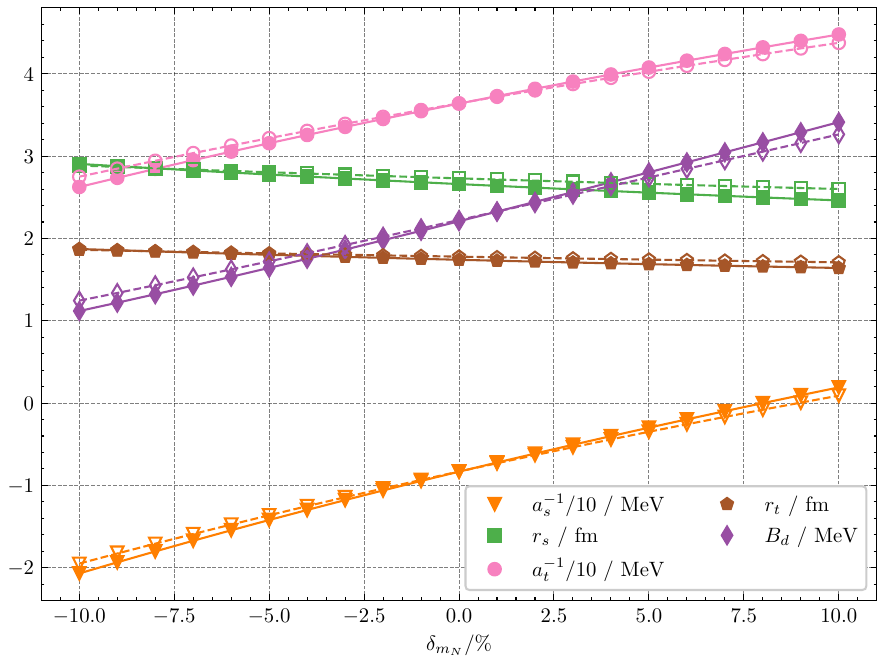} 
  \caption{Variation of the $np$ effective range expansion parameters with the nucleon mass: 
    $m_N(\delta_{m_N}) = m_N\,(1+\delta_{m_N})$, where $m_N$ 
    is the nominal nucleon mass and $\delta_{m_N}$ is the fractional change; 
    $a_s$ is the singlet scattering length (orange),
    $r_s$ is the corresponding effective range (green), 
    $a_t$ is the triplet scattering length (pink),
    $r_t$ is the corresponding effective range (brown), 
    $B_d$ is the deuteron binding energy (purple). 
    The closed symbols and solid lines refer to the NLEFT calculation of the nucleon-nucleon scattering parameters and the corresponding cubic polynomial fits, while the open symbols and dashed lines present the results obtained with the square-well potential and the corresponding fits.
  }
  \label{fig:scatBH}
\end{figure}

The variation of the effective range parameters is illustrated in Fig.~\ref{fig:scatBH}. We have fitted a cubic polynomial like
\begin{equation}\label{eq:cubic}
    X(\delta_{m_N}) = X_0 \cdot \left(1 + a \delta_{m_N}^{} + b \delta_{m_N}^2 + c \delta_{m_N}^3\right) 
\end{equation}
to the nucleon-nucleon scattering parameters. The fit parameters and corresponding errors are listed in Tab.~\ref{tab:NNparameters}. 
Note that we ignored the systematic errors in both calculations as we estimate the systematic error by the difference between the two methods.

\begin{table}[]
    \centering
    \begin{tabular}{c|c|c|c|c} \toprule
    
         \multicolumn{5}{c}{NLEFT}  \\ \midrule\midrule
        &  $X_0$ & $a$ & $b$ & $c$ \\ \midrule
        $a_s^{-1}$ & \SI{-8.3286 \pm 0.0004}{\mega\electronvolt} & \num{-13.437 \pm 0.001}& \num{12.802 \pm 0.009} & \num{-12.910 \pm 0.163} \\ 
        $r_s$ & \SI{2.66053 \pm 0.00001}{\femto\m} & \num{-0.8247 \pm 0.0001}& \num{0.872 \pm 0.001} & \num{-0.855 \pm 0.015} \\ 
        $a_t^{-1}$ & \SI{36.3849 \pm 0.0001}{\mega\electronvolt} & \num{2.5207 \pm 0.0001}& \num{-2.379 \pm 0.002} & \num{2.563 \pm 0.032} \\ 
        $r_t$ & \SI{1.740921 \pm 0.000001}{\femto\m} & \num{-0.65277 \pm 0.00001}& \num{0.905 \pm 0.001} & \num{-0.931 \pm 0.012} \\ 
        $B_d$ & \SI{2.20896 \pm 0.00009}{\mega\electronvolt} & \num{5.276 \pm 0.001}& \num{2.568 \pm 0.008} & \num{-8.240 \pm 0.158} \\ \midrule\midrule
        
        \multicolumn{5}{c}{Square-well potential}  \\ \midrule\midrule
        
        $a_s^{-1}$ & \SI{-8.310 \pm 0.008}{\mega\electronvolt} & \num{-12.182 \pm 0.030}& \num{11.840 \pm 0.212} & \num{-8.963 \pm 3.852} \\ 
        $r_s$ & \SI{2.7300 \pm 0.0002}{\femto\m} & \num{-0.519 \pm 0.001}& \num{0.498 \pm 0.009} & \num{-0.377 \pm 0.164} \\ 
        $a_t^{-1}$ & \SI{36.4147 \pm 0.0003}{\mega\electronvolt} & \num{2.2105 \pm 0.0002}& \num{-2.141 \pm 0.001} & \num{2.165 \pm 0.027} \\ 
        $r_t$ & \SI{1.778992 \pm 0.000001}{\femto\m} & \num{-0.42376 \pm 0.00004}& \num{0.3949 \pm 0.0003} & \num{-0.403 \pm 0.005} \\ 
        $B_d$ & \SI{2.22422 \pm 0.00007}{\mega\electronvolt} & \num{4.597 \pm 0.001}& \num{1.347 \pm 0.006} & \num{-5.733 \pm 0.117} \\ \bottomrule
    \end{tabular}
    \caption{Parameters for cubic polynomial fits, see Eq.~\ref{eq:cubic}, to nucleon-nucleon scattering observables calculated in NLEFT or using a square-well potential. Empirical values corresponding to the parameters listed under $X_0$ can be found in Tabs.~\ref{tab:par3S1}, \ref{tab:par1S0} of Appendix~\ref{app:square}.}
    \label{tab:NNparameters}
\end{table}

It is worth noting that the inverse singlet scattering length changes sign for $\delta_{m_N} \simeq 9\%$ and the
slope of the variation in the triplet scattering length is about half of the slope of the deuteron binding 
energy, consistent with the leading order result $B_d^{} = 1/(m_N^{} a_t^2)$. The changes in the effective range parameters
are more modest, as one would also expect on dimensional grounds.

\begin{figure}[!htb]
  \centering
  \includegraphics{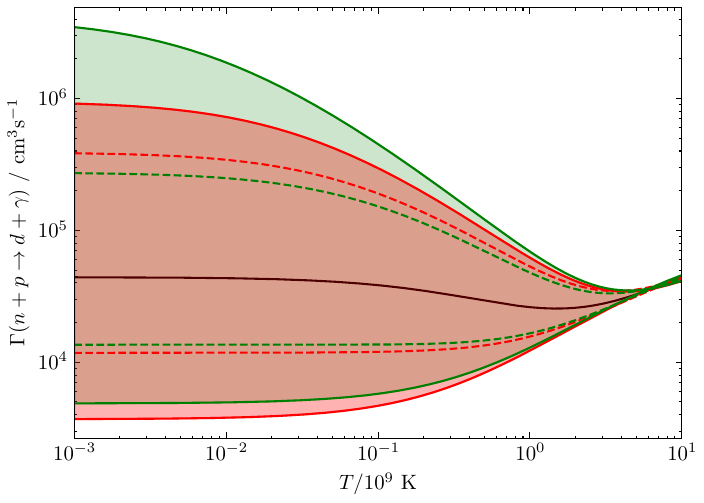} 
  \caption{Variation of the rate of the $n+p\to d + \gamma$ reaction as a function of the temperature $T_9 = T / (10^9\textnormal{ K})$.
  In red the variation based on the effective range parameters 
  calculated in the NLEFT framework is shown. The green curves are the results obtained with the simple square well potential. The black curve reflects the rate calculated with the nominal mass $m_N$, the dashed lines correspond to $\delta_{m_N}=\pm 0.05$ and the solid lines to $\delta_{m_N}=\pm 0.10$.}
  \label{fig:npdgmvar}
\end{figure}

The effect of changes in the nucleon mass on the $n+p \to d + \gamma$ rate as a function of the temperature is appreciable, 
see Fig.~\ref{fig:npdgmvar}, in particular for temperatures well below $T_9  = 1$, with $T_9 := T / (10^9\textnormal{ K})$.  However, the $n+p\to d+\gamma$ rate is mostly relevant at high temperatures so the effect from variations in this rate on the element abundances may not be as great as suggested by the figure. The main effect from changes in the deuteron binding energy will come from the inverse rate $d+\gamma \to n + p$ as this defines the start of BBN (the deuterium bottleneck).
 

No detailed theoretical description is available for other nuclear
reactions in the BBN network.  However, we can make a statement
concerning the variation of the binding energies of three- and four
nucleon systems with a variation of the singlet scattering length
$a_s$ and the deuteron binding energy $B_d$.  To this end we invoke
the relations cited in Eq.(5.2) in Ref.~\cite{Berengut:2013nh} quoting
\cite{Bedaque:2010hr}:
\begin{eqnarray}
\label{eq:KBPL}
  K^{a_s}_{B_{{{}^3}\textnormal{He}}}
  = 0.12 \pm 0.01\,,
  &&
     K^{B_d}_{B_{{{}^3}\textnormal{He}}}
     =
     1.41 \pm 0.01\,;
 \nonumber \\
  K^{a_s}_{B_{{{}^4}\textnormal{He}}}
  =
  0.037 \pm 0.011\,,
  &&
     K^{B_d}_{B_{{{}^4}\textnormal{He}}}
     =
     0.74 \pm 0.22\,,
\end{eqnarray}
where
\begin{equation}
\label{eq:Kxy}
  K^x_y := \frac{x}{y}\frac{\Delta y}{\Delta x}\,.
\end{equation}
From the parameter fits listed in Tab.~\ref{tab:NNparameters} and displayed in Fig.~\ref{fig:scatBH} we find for the NLEFT calculation
\begin{equation}
\label{eq:KNLEFT}
  K^{m_N}_{a_s}
  =
  -K^{m_N}_{1/a_s}
  =
  13.437 \pm 0.001\,,
  \qquad
  K^{m_N}_{B_d}
  =
  5.276 \pm 0.001\,,
\end{equation}
and from the results obtained with the square-well potential
\begin{equation}
\label{eq:KSQW}
  K^{m_N}_{a_s}
  =
  -K^{m_N}_{1/a_s}
  =
  12.18 \pm 0.03\,,
  \qquad
  K^{m_N}_{B_d}
  =
  4.597 \pm 0.001\,.
\end{equation}
We then obtain
\begin{equation}\label{eq:KEFT}
    \begin{aligned}
  K^{m_N}_{B_{{{}^3}\textnormal{He}}}
  &=&
      K^{m_N}_{a_s}\,
      K^{a_s}_{B_{{{}^3}\textnormal{He}}}
      +
      K^{m_N}_{B_d}\,
      K^{B_d}_{B_{{{}^3}\textnormal{He}}}
  =
      9.05 \pm 0.14\,,
  \\
  K^{m_N}_{B_{{{}^4}\textnormal{He}}}
  &=&
      K^{m_N}_{a_s}\,
      K^{a_s}_{B_{{{}^4}\textnormal{He}}}
      +
      K^{m_N}_{B_d}\,
      K^{B_d}_{B_{{{}^4}\textnormal{He}}}
  =
      4.4 \pm 1.2\,,
\end{aligned}
\end{equation}
from the NLEFT $np$-parameters and for the square-well potential calculation
\begin{equation}\label{eq:KSQW}
    \begin{aligned}
  K^{m_N}_{B_{{{}^3}\textnormal{He}}}
  &=&
      K^{m_N}_{a_s}\,
      K^{a_s}_{B_{{{}^3}\textnormal{He}}}
      +
      K^{m_N}_{B_d}\,
      K^{B_d}_{B_{{{}^3}\textnormal{He}}}
  =
      7.94 \pm 0.13\,,
  \\
  K^{m_N}_{B_{{{}^4}\textnormal{He}}}
  &=&
      K^{m_N}_{a_s}\,
      K^{a_s}_{B_{{{}^4}\textnormal{He}}}
      +
      K^{m_N}_{B_d}\,
      K^{B_d}_{B_{{{}^4}\textnormal{He}}}
  =
      3.9 \pm 1.0\,.
\end{aligned}
\end{equation}

Although the resulting $K$-factors for $^3$He differ by about 10$\%$, in the case of $^4$He, the $K$-factors agree within the uncertainties. 
We note that these are approximate relations, as they ignore any effective range effects in the binding
energies of the light nuclei. Using these $K$-factors, we find for the change of the binding energy as a function of
the change in the nucleon mass in linear approximation
\begin{displaymath}
  \Delta B_{{{}^3}\textnormal{He}}
  =
  \frac{B_{{{}^3}\textnormal{He}}}{m_N}\,K^{m_N}_{B_{{{}^3}\textnormal{He}}}\,\Delta m_N
\end{displaymath}
and thus
\begin{displaymath}
  B_{{{}^3}\textnormal{He}}(\delta_{m_N})
  =
  B_{{{}^3}\textnormal{He}}
  \left(1+K^{m_N}_{B_{{{}^3}\textnormal{He}}}\,\delta_{m_N}\right)\,,
\end{displaymath}
where $\delta_{m_N} := \Delta m_N/m_N$ is the fractional change in the nucleon mass.
Likewise
\begin{displaymath}
  B_{{{}^4}\textnormal{He}}(\delta_{m_N})
  =
  B_{{{}^4}\textnormal{He}}
  \left(1+K^{m_N}_{B_{{{}^4}\textnormal{He}}}\,\delta_{m_N}\right)\,.
\end{displaymath}
In both cases the dependence via the deuteron binding energy dominates.

\begin{figure}
    \centering
    \includegraphics[width=0.49\textwidth]{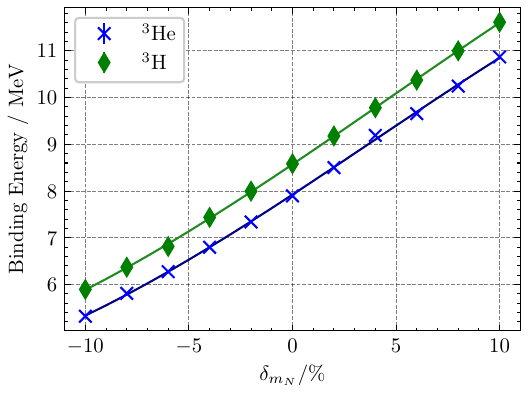}
    \includegraphics[width=0.49\textwidth]{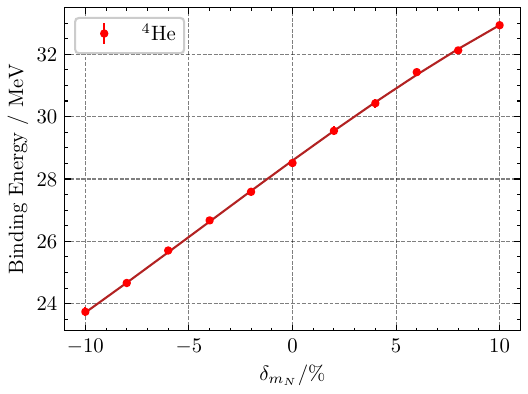}
    \caption{The binding energies for the three-nucleon system (left) and Helium-4 (right) calculated in the framework of NLEFT for a variation of the nucleon mass, $\delta_{m_N}$, between $\pm 10\%$. The solid lines are the corresponding cubic polynomial fits to the data, as given in Eq.~\eqref{eq:cubic}.}
    \label{fig:BE_3and4}
\end{figure}

Alternatively, we calculated the variation of the binding energies of the
three and four nucleon system within NLEFT. In both cases the same action as described in \cite{Elhatisari:2022zrb} is used, and the binding energies are calculated on a lattice with $L=10$ (in lattice units) in each spatial dimension and a lattice spacing $a=1.32\,$fm. For the three nucleon system, we calculate the binding energies non-perturbatively, while for the binding energy of Helium-4 we do various simulations for different values of the Euclidean time (for more details on Euclidean time projection, see, e.g., \cite{Lahde:2019npb}) so that we can extrapolate to infinite times. As for the calculation of the deuteron binding energy variation, we only varied the nucleon mass in the kinetic energy, the interactions in the Hamiltonian were not changed. 
The results are displayed in Fig.~\ref{fig:BE_3and4}.  

\begin{table}[]
    \centering
    \begin{tabular}{c|c|c|c|c} \toprule
        &  $X_0$ / MeV & $a$ & $b$ & $c$ \\ \midrule
        ${}^3\mathrm{H}$ & \num{8.571 \pm 0.012} & \num{3.477 \pm 0.039}& \num{2.011 \pm 0.272} & \num{-14.490 \pm 5.034} \\ 
        ${}^3\mathrm{He}$ & \num{7.923 \pm 0.019} & \num{3.662 \pm 0.066}& \num{2.152 \pm 0.458} & \num{-18.313 \pm 8.460} \\ 
        ${}^4\mathrm{He}$ & \num{28.591 \pm 0.028} & \num{1.685 \pm 0.025}& \num{-0.930 \pm 0.170} & \num{-7.367 \pm 3.453} \\ \bottomrule
    \end{tabular}
    \caption{Parameters for cubic polynomial fits (see Eq.~\eqref{eq:cubic}) to the three-nucleon system and Helium-4 binding energies calculated in the framework of NLEFT.}
    \label{tab:BE}
\end{table}

Similar to the nucleon-nucleon scattering observables, we have fitted a cubic polynomial (see Eq.~\eqref{eq:cubic}) to the binding energy data obtained in the NLEFT framework. The corresponding fit parameters are listed in Tab.~\ref{tab:BE}. The parameter of the term linear in $\delta_{m_N}$ is then the $K$-value defined in Eq.~\eqref{eq:Kxy}) as introduced e.g. in~\cite{Bedaque:2010hr}. A comparison of the results obtained within this framework:
\begin{equation}
\label{eq:K34NLEFT}
    K_{B_{{}^3\mathrm{H}}}^{m_N} = \num{3.477 \pm 0.039}\qc K_{B_{{}^3\mathrm{He}}}^{m_N} = \num{3.662 \pm 0.066}\qc K_{B_{{}^4\mathrm{He}}}^{m_N} = \num{1.685 \pm 0.025}\qc
\end{equation}
with the $K$-factors quoted in Eqs.~\eqref{eq:KEFT}-\eqref{eq:KSQW} shows that the dependence on $\delta_{m_N}$ is significantly smaller than what was found on the basis of the results from \cite{Berengut:2013nh} and \cite{Bedaque:2010hr}, 
especially for the three-nucleon system. We have ignored the systematic uncertainties in the NLEFT calculations (truncation error of perturbative expansion, lattice spacing dependence etc.), because as we can see, the systematic uncertainty from modeling the nucleon mass dependence as discussed here is considered to be more significant. We stress again that while the $K$-factors from the NLEFT calculation are exact (modulo higher order corrections),
the ones based on pionless EFT are subject to effective range corrections and thus turn out to be larger in size.

These results are then used to calculate the variation of
the $Q$-values for some relevant reactions involving these nuclei. The 
$Q$-values in turn affect the kinematics of the reactions, the reaction rate, the rates
of the inverse reactions as well as Coulomb penetration factors as
will be explained below.

\subsection{Modeling the nucleon mass dependence of nuclear reaction rates}
Based on the variation of the masses $m_i$ (or, equivalently, of the binding energies $B_i$) of the nuclei $\nuclide{2}{H}, \nuclide{3}{H}, \nuclide{3}{He}$ and $\nuclide{4}{He}$ with the nucleon mass, 
we modeled the variation of the rates of reactions involving these nuclei similar to the procedure we outlined in a previous paper on the variation with the electromagnetic fine structure constant, see ref.~\cite{Meissner:2023voo} for more details:

First of all a variation of masses in a reaction of the form $a + b \to c + d$ trivially leads to a variation of the reduced masses $\mu_{ab} = m_a\,m_b/(m_a+m_b)$
and $\mu_{cb} = m_c\,m_d/(m_c+m_d)$ in the entrance and exit channels. In addition, 
the $Q$-value of the reaction
\begin{equation}
    \label{eq:Qvalue}
    Q := m_a + m_b - m_c - m_d = B_c + B_d - B_a - B_b
\end{equation}
is affected. These two quantities enter the nuclear reactions rates as follows:
\begin{itemize}
    \item 
    The temperature dependent reaction rate that follows from the energy dependent total cross section via
    \begin{equation}
        \label{eq:reactionrate}
        \Gamma_{ab\to cd}(T)
        =
        N_A\,\sqrt{\frac{8}{\pi\,\mu_{ab}(k_B T)^3}}\int_0^\infty\textrm{d}E\,E\,\sigma_{ab\to cd}(E)\,\exp\left[-\frac{E}{k_B T}\right]
    \end{equation}
    depends on $\mu_{ab}$. Here $E$ is the center-of mass kinetic energy, $N_A$ denotes Avogadro's number and $k_B$ is Boltzmann's constant;
    \item 
    The reaction rate for the inverse reaction then depends on the reduced masses and the $Q$-value as
    \begin{equation}
        \label{eq:inverserate}
        \Gamma_{cd \to ab}(T)
        =
        \left(\frac{\mu_{ab}}{\mu_{cd}}\right)^{\frac{3}{2}}\,
        \frac{g_a\,g_b}{g_c\,g_d}\,\exp\left[-\frac{Q}{k_B T}\right]
         \Gamma_{ab\to cd}(T)\,,
    \end{equation}
    where $g_i$ denotes the spin multiplicity of particle $i$;
    \item 

We shall assume that the cross section for a strong
reaction $a + b \to c + d$ depends on $m_N$ as
\begin{equation}
    \label{eq:csdepabcd}
\sigma_{ab \to cd}(E; m_N)
=
\sqrt{E+Q(m_N)}\,
P_i(x_i(E,m_N))\, P_f(x_f(E,m_N))\,f(E)
\end{equation}
where $f$ is some function independent of $m_N$ and $P_i(x_i)$,
$P_f(x_f)$ are penetration factors of the form
 \begin{equation}
        \label{eq:penetrationfactor}
        P(x) 
        =
        \frac{x}{\exp{x}-1}\,,
    \end{equation}
reflecting the Coulomb repulsion in a channel where both particles are charged. 
The first factor in equation~(\ref{eq:csdepabcd})
accounts for the exit channel momentum dependence of the
cross section of the strong reaction $a + b \to c + d$. Here,
\begin{equation}
    \label{eq:Parg}
    x_i(E,m_N) =
    \sqrt{\frac{E_G^i(m_N)}{E}}\,,
    \qquad
        x_f(E,m_N) =
    \sqrt{\frac{E_G^f(m_N)}{E+Q(E)}}\,,
\end{equation}
with, denoting the charge number of nuclide $k$ by $Z_k$ and the fine structure constant by $\alpha$,
\begin{equation}
    \label{GamovEnergies}
    E_G^i = 2\,\pi^2\,Z_a^2\,Z_b^2\,\mu_{ab}\,\alpha^2\,,
    \quad
    E_G^f = 2\,\pi^2\,Z_c^2\,Z_d^2\,\mu_{cd}\,\alpha^2
\end{equation}
the Gamow-energies in the entrance and the exit channel,
respectively. 
Thus the cross section is supposed to depend on the varying nucleon mass 
$m_N(\delta_{m_N})=m_N\,(1+\delta_{m_N})$, where $m_N$ is the nominal nucleon mass and $\delta_{m_N}$ the fractional change, as
\begin{eqnarray}
    \label{eq:cssdep}
    \sigma_{ab \to cd}(E; m_N(\delta_{m_N}))
    &=&
    \frac{P_i(x_i(E,m_N(\delta_{m_N}))}{P_i(x_i(E,m_N))}\,
    \frac{\sqrt{E+Q(m_N(\delta_{m_N}))}}{\sqrt{E+Q(m_N)}}\,
    \frac{P_f(x_f(E,m_N(\delta_{m_N}))}{P_f(x_f(E,m_N))}\,
    \nonumber\\
    &&\quad\times\quad \sigma_{ab \to cd}(E; m_N)\,.
\end{eqnarray}
For a radiative capture reaction $a+b \to c + \gamma$ the corresponding expression is 
\begin{equation}
    \label{eq:csrdep}
    \sigma_{ab \to c\gamma}(E; m_N(\delta_{m_N})) = 
    \frac{P_i(x_i(E,m_N(\delta_{m_N}))}{P_i(x_i(E,m_N))}\,
    \left(\frac{E+Q(m_N(\delta_{m_N}))}{E+Q(m_N)}\right)^3\,
    \sigma_{ab \to c\gamma}(E; m_N)\,,
\end{equation}
where the second factor reflects the final state momentum dependence assuming
dipole dominance of the radiation.
\end{itemize}

\begin{figure}
    \centering
    \includegraphics[width=0.95\linewidth]{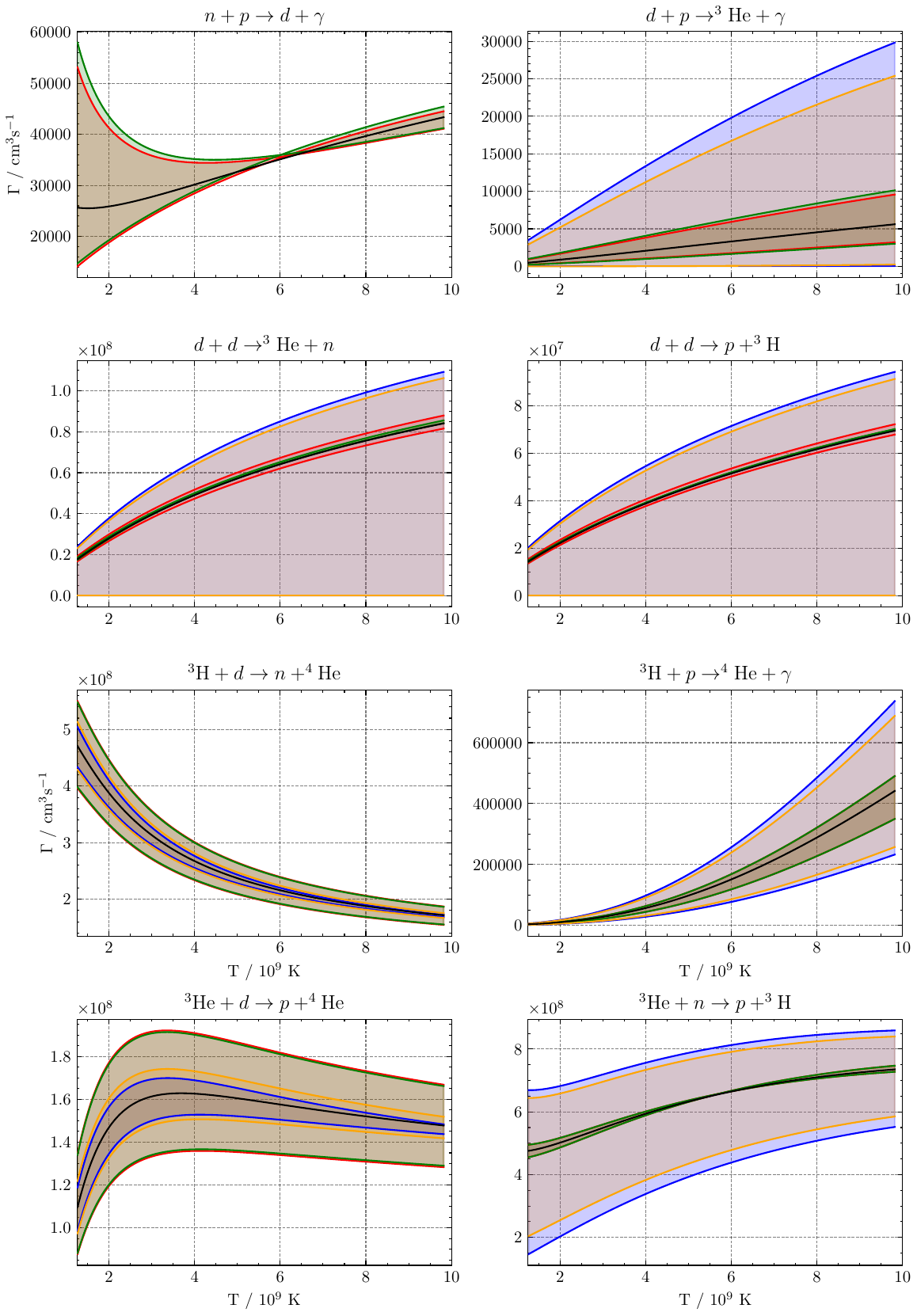}
    \caption{Reaction rates of the eight reactions in the BBN network involving only $d$, ${}^3\mathrm{H}$, ${}^3\mathrm{He}$ and ${}^4\mathrm{He}$. The nominal value of the rates according to \cite{Rupak:1999rk} (for $n+p \to d+\gamma$) or to the parameterizations listed in App.~\ref{app:params} is given in black. The bands correspond to a nucleon mass variation of $\pm 10\%$. Red: $(0,0)$; blue: $(0,1)$; green: $(1,0)$; orange: $(1,1)$, where the combinations $0$ and $1$ refer to the flags defined at the end of section \ref{sec:calc}.}
    \label{fig:rates}
\end{figure}

Considering exclusively the variations for the reactions including only $d$, ${}^3\mathrm{H}$, ${}^3\mathrm{He}$ and ${}^4\mathrm{He}$ is justified as these are the reactions most relevant for the abundances of these light nuclei, as was shown, e.g., in \cite{Serpico:2004gx}. In order to derive constraints on the nucleon mass variation, we will use experimental data collected on the abundances by the Particle Data Group (PDG) \cite{ParticleDataGroup:2024cfk} for deuterium and Helium-4, because these are the most reliable (for Lithium-7 the so-called \enquote{Lithium problem} exists~\cite{Fields:2011zzb},  which makes it unsuitable for these kinds of considerations). The binding energy dependence on the nucleon mass for other light nuclei is not known, so it would seem incomplete (and unnecessary) to change the $Q$-values of other reactions that include the four lightest but also other nuclei. 
Including the first reaction $n+p\to d+\gamma$ which was already discussed in Sect.~\ref{ss:BE}, there are eight reactions where only the four lightest nuclei are involved. In Fig.~\ref{fig:rates}, the variation of these rates resulting from changing the nucleon mass by $\pm 10\%$ is displayed for the temperature range in which these reactions are most relevant in the BBN process as explained in \cite{PRIMAT-companion}. 
Here, we have used an updated version (compared to \cite{Meissner:2023voo}) of the parameterizations of cross sections for 17 key rates which are listed in Appendix~\ref{app:params}.

We have defined \enquote{flags} corresponding to different combinations of how we calculated the nucleon-nucleon scattering parameters and the changes in the binding energies:
\begin{itemize}
    \item ($0,0$) means everything is based on NLEFT calculations (see the upper part of Tab.~\ref{tab:NNparameters} and Tab.~\ref{tab:BE}); 
    \item ($0,1$) is the NLEFT result for the nucleon-nucleon parameters (see the upper part of Tab.~\ref{tab:NNparameters}) combined with variations of the binding energies from  pionless EFT~\cite{Bedaque:2010hr} (see Eq.~\eqref{eq:KEFT});
    \item  ($1,0$) uses the square-well results for the $np$ scattering parameters  (see the lower part of Tab.~\ref{tab:NNparameters}) combined with the three- and four-body binding energies from NLEFT (Tab.~\ref{tab:BE});
    \item ($1,1$) is again the $np$ scattering parameters (see the lower part of Tab.~\ref{tab:NNparameters}) from the square-well potential now combined with the $K$-factors from pionless EFT~\cite{Bedaque:2010hr} (see Eq.~\eqref{eq:KSQW}).
\end{itemize}  
For some of these combinations, the $Q$-values for some reactions become negative for certain $\delta_{m_N}$ meaning the reactions is not kinematically allowed anymore. In this case, we have simply set the reaction rate to zero.


\section{Results and discussion}
\label{sec:res}

\begin{figure}
    \centering
    \includegraphics[width=\textwidth]{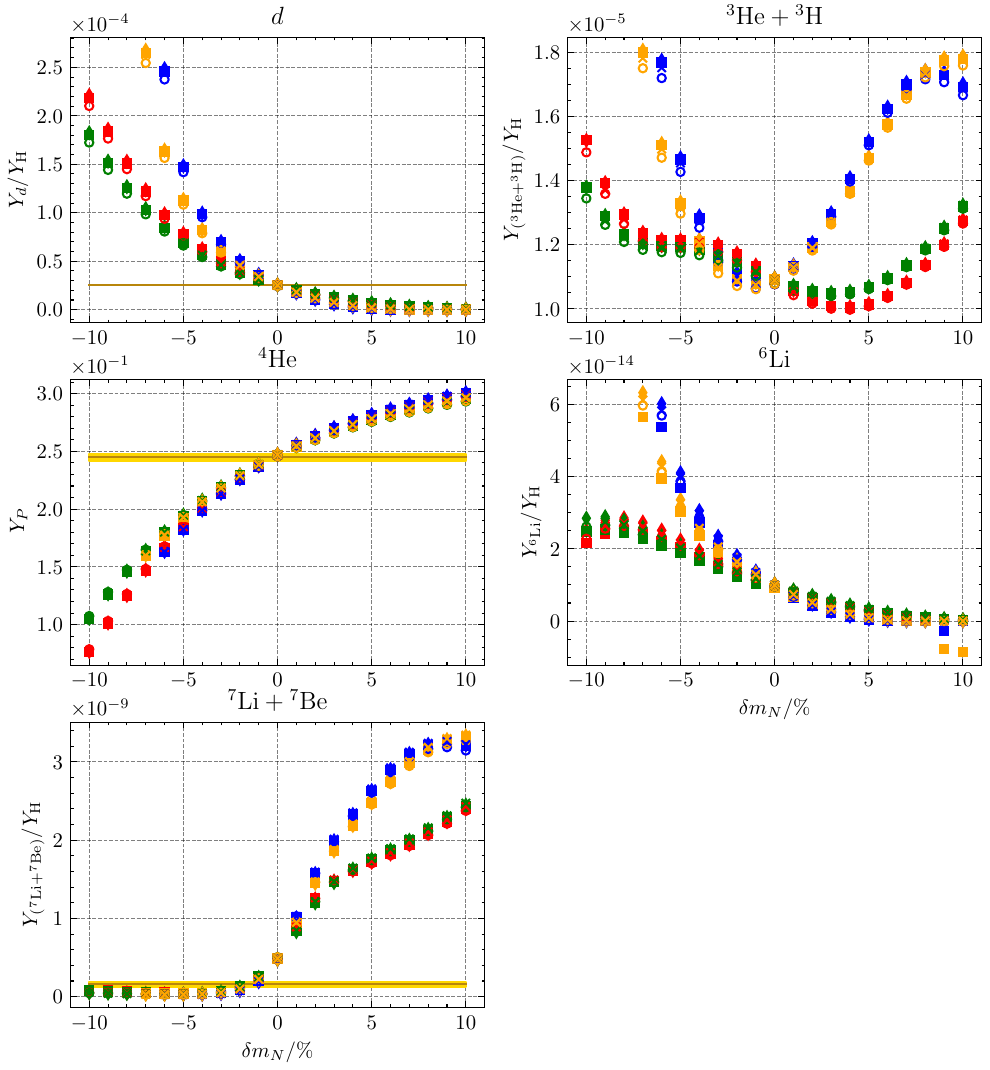}
    \caption{Nucleon mass dependence 
    of the relative abundances for ${}^2\textnormal{H}$, ${}^3\textnormal{H}+\textnormal{He}$,
    ${}^4\textnormal{He}$, ${}^6\textnormal{Li}$ and ${}^7\textnormal{Li}+\textnormal{Be}$ calculated with a modified verion of the Kawano code \texttt{nuc123} \cite{FERMILAB-PUB-92-004-A} (open circles), with \texttt{PRyMordial} \cite{burns2023prymordial} (squares), with \texttt{AlterBBN} \cite{Arbey:2011nf,Arbey:2018zfh} (diamonds) and with \texttt{PRIMAT} \cite{PRIMAT-companion} (crosses). For ${}^4\textnormal{He}$ this is the mass ratio with respect to hydrogen, for the other nuclides the relative abundence reflects the number ratio with respect to hydrogen. The color coding of the curves, respective to the different methods, is the same as in Fig.~\ref{fig:rates}.
    }
    \label{fig:abundances}
\end{figure}

The variation of the resulting relative abundances for the nuclides ${}^2\textnormal{H}$, ${}^3\textnormal{H}+{}^3\textnormal{He}$,
    ${}^4\textnormal{He}$, ${}^6\textnormal{Li}$ and ${}^7\textnormal{Li}+{}^7\textnormal{Be}$ with a variation of the nucleon mass,
    cf. Eq.~\eqref{eq:deltamN} is displayed in Fig.~\ref{fig:abundances}.  
    These results were obtained by varying the rates of the eight reactions discussed above with modified (see~\cite{Meissner:2022dhg,Meissner:2023voo}) versions of 
    \begin{itemize}
        \item 
         the Kawano code~\cite{FERMILAB-PUB-92-004-A} \texttt{nuc123},
        \item 
        \texttt{PRyMordial}~\cite{burns2023prymordial}, 
        \item 
        \texttt{AlterBBN}~\cite{Arbey:2011nf,Arbey:2018zfh} and 
        \item
        \texttt{PRIMAT} \cite{PRIMAT-companion}.
    \end{itemize}
    
    The nominal abundances, i.e. without variation of the nucleon mass are compared to experimental data in Tab.~\ref{tab:abundances_mN0}. 

      \begin{table}[htb!]
        \centering
        \begin{tabular}{c|c|c|c|c|c} \toprule
             & $Y_d / Y_H$ & $Y_{{}^3\mathrm{H}+{}^3\mathrm{He}} / Y_H$ & $Y_P$ & $Y_{{}^6\mathrm{Li}} / Y_H$ & $Y_{{}^7\mathrm{Li} + {}^7\mathrm{Be}} / Y_H$ \\ \midrule
            \texttt{nuc123} & \num{2.452 e-5} & \num{1.078 e-5}& \num{0.246} & \num{0.960 e-14}& \num{5.035 e-10} \\
            \texttt{PRyMordial} & \num{2.541 e-5} & \num{1.092 e-5} & \num{0.247} & \num{0.920 e-14}&\num{4.874 e-10} \\
            \texttt{AlterBBN} & \num{2.563 e-5} & \num{1.096 e-5} & \num{0.247} & \num{1.027 e-14}& \num{4.756 e-10}  \\
            \texttt{PRIMAT} &  \num{2.517 e-5} & \num{1.089 e-5} & \num{0.247} & \num{0.994 e-14} & \num{4.874 e-10} \\ \midrule
            PDG \cite{ParticleDataGroup:2024cfk} & \num{2.547 \pm 0.029 e-5} & & \num{0.245 \pm 0.003} & & \num{1.6 \pm 0.3 e-10} \\ \bottomrule
        \end{tabular}
        \caption{Values of the light element abundances as calculated with the four codes for the nominal value of the nucleon mass $m_N$ and the corresponding experimental value from \cite{ParticleDataGroup:2024cfk}. For ${}^4\mathrm{He}$ the mass fraction is given.}
        \label{tab:abundances_mN0}
    \end{table}
    
\begin{figure}
        \centering
        \includegraphics[width=0.8\linewidth]{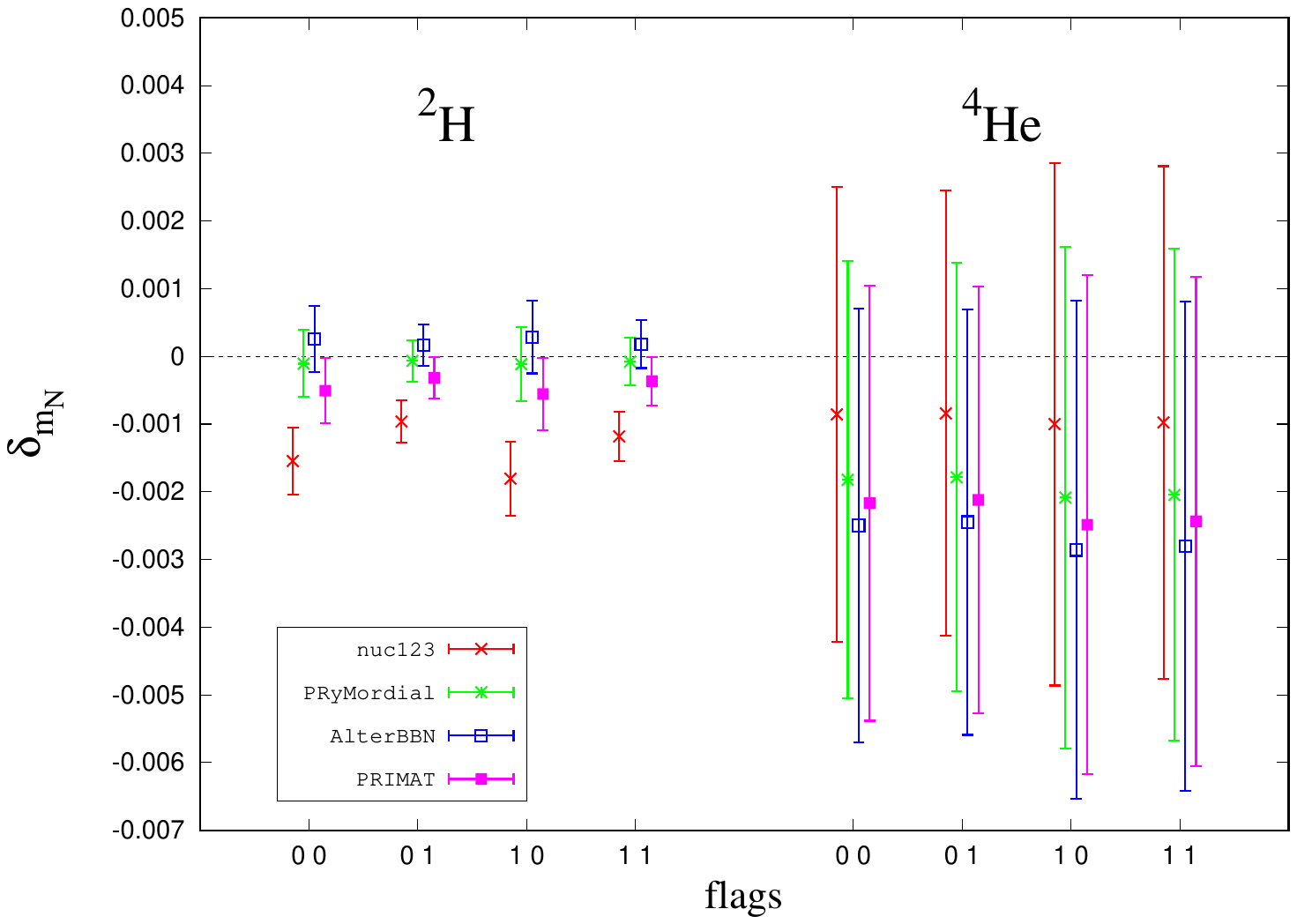}
        \caption{Extracted ranges of $\delta_{m_N}$ for the four rate combinations labeled "0 0", "0 1", "1 0" and "1 1" discussed in section~\ref{sec:calc}, both on the basis of the ${{}^2}$H (left) and the ${{}^4}$He (right) abundance data. The results with the BBN code \texttt{nuc123} is displayed by red crosses, with \texttt{PRyMordial} by green stars, with \texttt{AlterBBN} with blue open squares and with \texttt{PRIMAT} by magenta filled squares.}
        \label{fig:mnint}
    \end{figure}

    On the basis of the uncertainties of the empirical data for the relative deuteron abundance, $Y_{{{}^2}\textnormal{H}} = (2.547 \pm 0.029)\times 10^{-5}$\,\cite{ParticleDataGroup:2024cfk}, and the relative ${{}^4\textnormal{He}}$ (mass) abundance, $Y_{{{}^4}\textnormal{He}} = 0.245 \pm 0.003$\,\cite{ParticleDataGroup:2024cfk}, we estimated the possible variation of the nucleon mass with the four combinations of the rates discussed in Sect.~\ref{sec:calc}. The resulting ranges in the relative nucleon mass variation $\delta_{m_N}$ are displayed in Fig.~\ref{fig:mnint} 
    and listed in Tab.~\ref{tab:constraints_d} for ${}^2$H and in Tab.~\ref{tab:constraints_He4} for ${}^4$He.

    \begin{table}[htb!]
        \centering
        \begin{tabular}{cc|c|c|c|c|c|c|c|c} \toprule
             & & \multicolumn{2}{c|}{\texttt{nuc123}} & \multicolumn{2}{c|}{\texttt{PRyMordial}} & \multicolumn{2}{c|}{\texttt{AlterBBN}} & \multicolumn{2}{c}{\texttt{PRIMAT}}\\ \cmidrule{3-10}
            \multicolumn{2}{c|}{Flag} & $\delta_{m_N,\mathrm{min}}$ & $\delta_{m_N,\mathrm{max}}$ & $\delta_{m_N,\mathrm{min}}$ & $\delta_{m_N,\mathrm{max}}$ & $\delta_{m_N,\mathrm{min}}$ & $\delta_{m_N,\mathrm{max}}$ & $\delta_{m_N,\mathrm{min}}$ & $\delta_{m_N,\mathrm{max}}$\\ \midrule
            0 & 0 & -0.00204 & -0.00105 & -0.00060 & 0.00039 & -0.00023 & 0.00075 & -0.00099 & -0.00002\\
            0 & 1 & -0.00127 & -0.00065 & -0.00037 & 0.00024 & -0.00014 & 0.00047 & -0.00062 & -0.00001\\
            1 & 0 & -0.00235 & -0.00126 & -0.00066 & 0.00043 & -0.00025 & 0.00082 & -0.00109 & -0.00002\\
            1 & 1 & -0.00154 & -0.00082 & -0.00043 & 0.00028 & -0.00017 & 0.00054 & -0.00072 & -0.00001\\ \bottomrule
        \end{tabular}
        \caption{Constraints on the nucleon mass variation $\delta_{m_N}$ from comparing the simulated deuterium abundance for different values of $m_N$ to the experimental value $\num{2.547 \pm 0.029 e-5}$ \cite{ParticleDataGroup:2024cfk}, for the different methods of deriving the nucleon-nucleon scattering parameters and binding energies and for the different codes used to simulate BBN.}
        \label{tab:constraints_d}
    \end{table}

    \begin{table}[htb!]
        \centering
        \begin{tabular}{cc|c|c|c|c|c|c|c|c} \toprule
             & & \multicolumn{2}{c|}{\texttt{nuc123}} & \multicolumn{2}{c|}{\texttt{PRyMordial}} & \multicolumn{2}{c|}{\texttt{AlterBBN}} & \multicolumn{2}{c}{\texttt{PRIMAT}}\\ \cmidrule{3-10}
            \multicolumn{2}{c|}{Flag} & $\delta_{m_N,\mathrm{min}}$ & $\delta_{m_N,\mathrm{max}}$ & $\delta_{m_N,\mathrm{min}}$ & $\delta_{m_N,\mathrm{max}}$ & $\delta_{m_N,\mathrm{min}}$ & $\delta_{m_N,\mathrm{max}}$ & $\delta_{m_N,\mathrm{min}}$ & $\delta_{m_N,\mathrm{max}}$\\ \midrule
            0 & 0 & -0.00422 & 0.00251 & -0.00505 & 0.00141 & -0.00570 & 0.00071 & -0.00538 & 0.00105\\
            0 & 1 & -0.00413 & 0.00245 & -0.00495 & 0.00138 & -0.00559 & 0.00070 & -0.00527 & 0.00103\\
            1 & 0 & -0.00486 & 0.00286 & -0.00579 & 0.00162 & -0.00654 & 0.00082 & -0.00617 & 0.00120\\ 
            1 & 1 & -0.00476 & 0.00281 & -0.00568 & 0.00159 & -0.00641 & 0.00081 & -0.00605 & 0.00118\\ \bottomrule          
        \end{tabular}
        \caption{Constraints on the nucleon mass variation $\delta_{m_N}$ from comparing the simulated Helium-4 mass fraction for different values of $m_N$ to the experimental value $\num{0.245 \pm 0.003}$ \cite{ParticleDataGroup:2024cfk}, for the different methods of deriving the nucleon-nucleon scattering parameters and binding energies and for the different codes used to simulate BBN.}
        \label{tab:constraints_He4}
    \end{table}
   
 From this we infer that $\delta_{m_N} \in (-0.0024,+0.0008)$ and  $\delta_{m_N} \in (-0.0065,+0.0029)$, on the basis of $Y_{{{}^2}\textnormal{H}}$ and $Y_{{{}^4}\textnormal{He}}$, respectively. This is all well below $1\%$\,. It is amusing to observe, see the bottom panel in Fig.~\ref{fig:abundances}, that a decrease in the nucleon mass by $1-2\%$ apparently would solve the so called "Lithium-problem", but such a variation is of course excluded by the other relative abundance data mentioned before. For the abundances of  ${}^6\textnormal{Li}$ and ${}^3\textnormal{He}+{}^3\textnormal{H}$ a strong non-linear dependence on the nucleon mass is found, while also the four versions of the rates yield markedly varying results for the abundances, but unfortunately there are no reliable empirical data for these nuclides.



We can now draw conclusion on the size of the variation of the strangeness content and the strange quark mass. Taking the largest possible variation (in size) from Tabs.~\ref{tab:constraints_d},\ref{tab:constraints_He4}, we see that $|\Delta {m_N}|$
is bounded by $2.3$~MeV and $6.1$~MeV from the deuterium and the $^4$He abundance, respectively.
This translates into possible variations of $\sigma_s$ of less than $5.1\%$ and $13.6\%$, respectively, 
using the central FLAG value for $N_f =2+1$ flavors.  As in the case of variations 
of the Higgs VEV~\cite{Meyer:2024auq}, the deuterium abundance sets the stronger bound. Assuming that the strange quark condensate
does not change, we can derive an upper bound on variation of the strange quark mass (from the deuterium abundance)
\begin{equation}
\left| \frac{\Delta m_s}{m_s}\right| \leq 5.1\%~.
\end{equation}
Note, however, that this is only an upper bound. Scanning through Tabs.~\ref{tab:constraints_d},\ref{tab:constraints_He4}, one can see that the range of possible variations for the individual codes is much smaller. The interval of possible values for $\delta_{m_N}$ that we mention is only as large because the nominal value of the abundances vary between the codes (see also Fig.~\ref{fig:mnint}). In addition, if we would take the largest values of $\delta_{m_N}$ found from NLEFT calculations only (flag $(0,0)$), which, in our opinion, is the best calculation, the upper bound for the variation of the strange quark mass would reduce further by $1\%$.

\section*{Acknowledgments}
We thank Serdar Elhatisari and Fabian Hildenbrand for help with the NLEFT simulations. The authors gratefully acknowledge the computing time granted by the John von Neumann Institute for Computing (NIC) and provided on the supercomputer JURECA \cite{JURECA} at Jülich Supercomputing Centre (JSC).
This work was supported in part by the European
Research Council (ERC) under the European Union's Horizon 2020 research
and innovation programme through the ERC AdG EXOTIC (grant agreement No. 101018170).
The work of UGM was supported in part by the CAS President's International
Fellowship Initiative (PIFI) (Grant No.~2025PD0022).

\appendix

\section{Square-well potentials} 
\label{app:square}

The stationary radial {\sc Schr\"odinger} equation for non-relativistic S-wave 
relative motion of two nucleons of mass $m_N$ 
in a spherical potential well of depth $-V_0$ and width $R_0$\,:
\begin{equation}
\label{eq:sqwp}
  V(r)
  =
  \left\lbrace
    \begin{array}[c]{rl}
      -V_0\,, & r<R_0\,,
      \\
      0\,, & r> R_0
    \end{array}
  \right.\,,
\end{equation}
reads 
\begin{equation}
\label{eq:sse}
  -\frac{1}{2\mu}\,u''(r) + V(r)\,u(r) = E\,u(r)
\end{equation}
where $\mu = m_N/2$ is the reduced mass and $E$ the energy.
Defining the dimensionless potential strength 
\begin{equation}  
\label{eq:xi}
\xi^2 := 2\,\mu\,V_0\,R_0^2\,,
\end{equation}  
the scattering length $a$ and the effective range $r$ 
are given by the expressions:
\begin{equation}
\label{eq:ar}
a= R_0\,\alpha(\xi) := R_0\left(1-\frac{\tan{\xi}}{\xi}\right)\,,
\qquad
r = R_0\left(1-\frac{1}{\alpha(\xi)\,\xi^2}-\frac{1}{3\,\alpha(\xi)^2}\right)\,.
\end{equation}
Accordingly, for given $a$ and $r$ the value of $\xi^2 \propto V_0\,R_0^2$ follows from the ratio $r/a$ and $R_0$ from the first equation in (\ref{eq:ar}).
For a bound state $\xi>1$ and with $x$ the solution of $\sin(x)/x=1/\xi$ the binding energy $B$ is given by
\begin{displaymath}
    B = \frac{1}{2\,\mu\,R_0^2}\left(\xi^2-x^2\right)\,.
\end{displaymath}

\paragraph{Application to the np-${}^3 S_1$ channel:}

In this simple model we consider S-wave scattering only and thus neglect
the ${}^3 S_1 - {}^3 D_1$ mixing. 
The low energy triplet scattering parameters (scattering length $a_t$ and effective range $r_t$) can \textit{e.g.} be found in Ref.~\cite{deSwart:1995ui}.
With the spherical well parameters listed in Tab.~\ref{tab:par3S1}.
\begin{table}[!htb]
  \centering
  \begin{tabular}{|l|l|c|}
    \hline
    proton mass & $m_p~[\textnormal{MeV}]$  & $938.27208816$ \\
    neutron mass & $m_n~[\textnormal{MeV}]$ & $939.56542052$ \\
    reduced mass & $\mu~[\textnormal{MeV}]$ & $469.45915448$ \\
    nucleon mass & $m_N~[\textnormal{MeV}]$ & $938.91875434$ \\
    \hline
    depth        & $V_0~[\textnormal{MeV}]$ & $33.306$ \\
    radius       & $R_0~[\textnormal{fm}]$ & $2.116$ \\
    \hline
  \end{tabular}
\caption{Spherical well parameters for the ${}^3 S_1$ channel.}
  \label{tab:par3S1}
\end{table}
we find with the values for the scattering length and effective range as listed in the column labeled "calc." in Tab.~\ref{tab:res}, a fair description of various empirical deuteron data, also listed in Tab.~\ref{tab:res}. 
\begin{table}[!htb]
  \centering
  \begin{tabular}{|l|l|c|c|}
  \hline
    & & calc. & exp.~\cite{deSwart:1995ui}. \\
    \hline
    inverse scattering length & $a_t^{-1}~[\textnormal{MeV}]$ & $36.414$    & $36.407(7)$ \\
    effective range   & $r_t~[\textnormal{fm}]$ & $1.779$    & $1.753(2)$ \\
    binding energy    & $B~[\textnormal{MeV}]$  & $2.224433$ & $2.224575(9)$ \\
    asymptotic normalisation & $A_s~[\textnormal{fm}^{-\frac{1}{2}}]$ & $0.879$ & $0.8845(8)$ \\
    root mean square radius  & $r_d~[\textnormal{fm}]$ & $1.9505$ & $1.9676(10)$ \\
    \hline
  \end{tabular}
   \caption{Deuteron properties and low-energy scattering parameters for the ${}^3 S_1$ channel.}
  \label{tab:res}
\end{table}

\paragraph{Application to the np-${}^1 S_0$ channel:}

We have inspected three parameter sets, based on three combinations of
$(a_s, r_s)$ (the ${}^1 S_0$ scattering length and effective range, respectively), as listed in Tab.~\ref{tab:par1S0}, that, via Eq.~\eqref{eq:ar} lead to the potential parameters $V_0$ and $R_0$ listed in the last two columns of this table.
\begin{table}[!htb]
  \centering
  \begin{tabular}[c]{|l|c|c|c|c|}
    \hline
    & $a_s^{-1}~\textnormal{[MeV]}$ & $r_s~\textnormal{[fm]}$ & $V_0~\textnormal{[MeV]}$ & $R_0~\textnormal{[fm]}$ \\
    \hline
    Experiment~\cite{Machleidt:2000ge} & -8.312(7) & 2.77(5) & 13.358 & 2.650 \\
    Pionless EFT~\cite{Rupak:1999rk} & -8.309 & 2.73 & 13.752 & 2.614 \\
    CD-Bonn~\cite{Machleidt:2000ge} & -8.313 & 2.671 & 14.376 & 2.558 \\
    \hline
  \end{tabular}
  \caption{Scattering parameters and spherical well parameters for the ${}^1 S_0$ channel.}
\label{tab:par1S0}   
\end{table}
Keeping the potential parameters fixed to the values of Tabs.~\ref{tab:par3S1},\ref{tab:par1S0} (second line labeled "Pionless EFT") by varying the nucleon mass as $m_N(\delta_{m_N}) = m_N\,(1+\delta_{m_N})$ the variation of the scattering parameters and the deuteron binding energy is as shown in Fig.~\ref{fig:scatBH}.

\section{Parameterizations of cross sections}\label{app:params}
In this appendix we present updated parameterizations for $17$ relevant cross sections in BBN. Compared to the parameterizations shown in \cite{Meissner:2023voo}, we have slightly changed a few parameters and added error estimations. Because of its more modest variation with the centre-of-mass (CMS) energy $E$, one often prefers to parameterize the so-called $S$-factor
\begin{equation}
	S(E) = \sigma(E)\cdot E \cdot e^{\sqrt{E_G / E}}
\end{equation}
over the cross section $\sigma(E)$. Here the Gamow-energy $E_G$ is 
\begin{equation}\label{E_G}
E_G = 2\pi^2 \mu_{ab}c^2 Z_a^2 Z_b^2 \alpha^2,
\end{equation}
and the $S$-factor was chosen to be a rational function in terms of the CMS energy $E$ (given here in \si{\mega\electronvolt}) multiplied by a constant $S_0$ (in \si{\mega\electronvolt \milli\barn}):
\begin{equation}\label{eq:SRatPol}
S(E) = S_0\, \frac{1 + a_1 E + a_2 E^2 + a_3 E^3}{1 + q_1 E + q_2 E^2 + q_3 E^3}.
\end{equation}
The constant $S_0$ and the coefficients $a_k, q_k$ in $(\si{\mega\electronvolt})^{-k}$ are given in Tabs.~\ref{tab:Par_RadCap} and \ref{tab:Par_Charged} for $14$ charged particle reactions.

\noindent
For neutron capture reactions we parameterize a function
\begin{equation}
\mathcal{R}(E) = \sigma(E)\sqrt{E}
\end{equation}
similarly to the $S$-factor above as
\begin{equation}\label{eq:R_NC}
\mathcal{R}(E) = \mathcal{R}_0 \frac{1 + a_1 E + a_2 E^2 + a_3 E^3}{1 + q_1 E + q_2 E^2 + q_3 E^3},
\end{equation}
with $E$ in \si{\mega\electronvolt}, the coefficients $a_k, q_k$ in $(\si{\mega\electronvolt})^{-k}$ and $\mathcal{R}_0$ in \si{(\mega\electronvolt)^{1/2}\milli\barn}. 
The $S$-factor is then for neutron-capture reactions
\begin{equation}
S(E) = \mathcal{R}(E)\cdot \sqrt{E} = \sigma(E)\cdot E.
\end{equation}

There are some reactions with resonances in the energy range considered here. For these, we can parameterize the $S$-factor as a rational function or a polynomial with a combination of (relativistic) Breit-Wigner functions. The coefficients for the relativistic Breit-Wigner functions ($E, \Gamma, M$ in \si{\mega\electronvolt}) of the form
\begin{equation}\label{eq:BWfunc}
	\mathrm{BW}(E ; k, \Gamma, M) = \frac{k}{\Gamma^2M^2 + (E^2 - M^2)^2}
\end{equation}
and the non-relativistic ones like
\begin{equation}\label{eq:BWfunc_nr}
\mathrm{bw}(E ; k, \kappa, M) = \frac{k}{1 + \kappa(E - M)^2},
\end{equation}
with $\kappa$ in $\si{(\mega\electronvolt)^{-2}}$ can be found in Tab.~\ref{tab:Par_BW} for the pertinent reactions.

For the function fits we use data composed by EXFOR \cite{Exfor_lib, Exfor_web} for the cross section of $17$ relevant reactions in BBN.

\subsection{Radiative capture reactions}\label{as:RadCap}

\begin{figure}
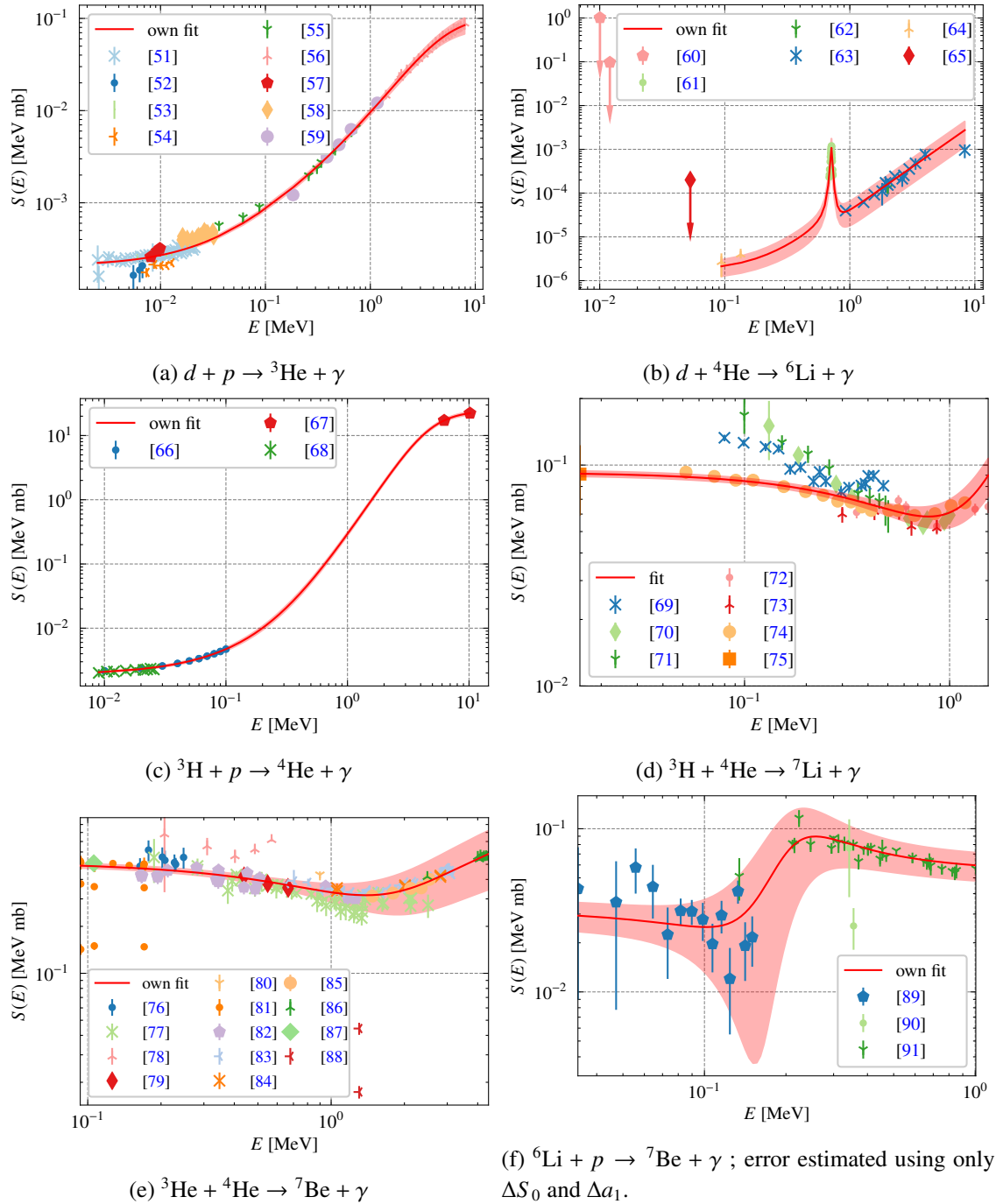

	\centering
	\begin{subfigure}{.49\textwidth}
		\resizebox{\textwidth}{!}{\input{figs/Params/dpgHe3_error.pgf}}
		\caption{$d + p \to \He{3} + \gamma$}
	\end{subfigure}
	\begin{subfigure}{.49\textwidth}
		\resizebox{\textwidth}{!}{\input{figs/Params/dagLi6_error.pgf}}
		\caption{$d + \He{4} \to \Li{6} + \gamma$}
	\end{subfigure}
	\begin{subfigure}{.49\textwidth}
		\resizebox{\textwidth}{!}{\input{figs/Params/H3pga_error.pgf}}
		\caption{$\triton + p \to \He{4} + \gamma$}
	\end{subfigure}
	\begin{subfigure}{.49\textwidth}
	\resizebox{\textwidth}{!}{\input{figs/Params/H3agLi7_error.pgf}}
	\caption{$\triton + \He{4} \to \Li{7} + \gamma$}
	\end{subfigure}
	\begin{subfigure}{.49\textwidth}
		\resizebox{\textwidth}{!}{\input{figs/Params/He3agBe7_error.pgf}}
		\caption{$\He{3} + \He{4} \to \Be{7} + \gamma$}
	\end{subfigure}
	\begin{subfigure}{.49\textwidth}
		\resizebox{\textwidth}{!}{\input{figs/Params/Li6pgBe7_error.pgf}}
		\caption{$\Li{6} + p \to \Be{7} + \gamma$ ; error estimated using only $\Delta S_0$ and $\Delta a_1$.}
	\end{subfigure}
	
	\caption{Fits to experimental data composed by EXFOR \cite{Exfor_web} with error band for some relevant radiative capture reactions. The arrows on the data points in panel~(b) indicate that these are upper bounds, not included in the fit.}
	\label{fig:radcap_par}
\end{figure}

The parameters found by fitting data composed by \cite{Exfor_web} to Eq.~\eqref{eq:SRatPol} are displayed in Tab.~\ref{tab:Par_RadCap} for most radiative capture reactions treated here. The parameterizations are compared to the corresponding data in Fig.~\ref{fig:radcap_par}.

\begin{table}[htb]
	\centering
	\sisetup{table-alignment=center, round-mode=places, round-precision=3, separate-uncertainty}
	\resizebox{\textwidth}{!}{%
	\begin{tabular}{@{}l%
			@{\extracolsep{1em}}S[table-format = 1.3e2, table-figures-uncertainty=1]%
			@{\extracolsep{0.5em}}S[table-format = 2.3, table-figures-uncertainty=1]%
			@{\extracolsep{0.5em}}S[table-format = 3.3, table-figures-uncertainty=1]%
			@{\extracolsep{0.5em}}S[table-format = 3.3, table-figures-uncertainty=1]%
			@{\extracolsep{0.5em}}S[table-format = 2.3, table-figures-uncertainty=1]%
			@{\extracolsep{0.5em}}S[table-format = 2.3, table-figures-uncertainty=1]%
			@{\extracolsep{0.5em}}S[table-format = 1.3e1, table-figures-uncertainty=1]%
			@{}}
		\toprule
		{Reaction} & {$S_0$} & {$a_1$} & {$a_2$} & {$a_3$} & {$q_1$} & {$q_2$} & {$q_3$}\\
		\midrule
		
		{$d + p \to \He{3} + \gamma$} & 2.072 \pm 0.043 e-4  & 30.094 \pm 1.839 & 16.754 \pm 2.302 & 0 & 0 & 0.035 \pm 0.008 & 0\\

		{$d + \He{4} \to \Li{6} + \gamma$} & 4.382 \pm 1.762 e-7 & 0 & 90.682 \pm 37.560 & 0 & 0 & 0 & 0\\
		
		{$\triton + p \to \He{4} + \gamma$} & 1.875 \pm 0.031e-3 & 10.773 \pm 0.996 & 32.613 \pm 7.711 & 113.840 \pm 4.597 & 0 & 0 & 8.919 \pm 0.181e-3\\
		
		{$\triton + \He{4} \to \Li{7} + \gamma$} & 9.275 \pm 0.258 e-2 & -0.937 \pm 0.067 & 0.594\pm 0.064 & 0 & 0 & 0&0 \\
	
		{$\He{3} + \He{4} \to \Be{7} + \gamma$} &5.160 \pm 0.161 e-1 & -0.556 \pm 0.064 & 0.281 \pm 0.060 & 0 & 0 & 0.132 \pm 0.051 &0\\
		
		{$\Li{6} + p \to \Be{7} + \gamma$} & 3.208 \pm 0.619 e-2 & 40.354 \pm 9.384 & -11.448 \pm 1.107 & 0 & -9.092 \pm 0.616 & 24.154 \pm 3.089 & 0 \\ \bottomrule
	\end{tabular}}
		\caption{  \label{tab:Par_RadCap}
		Fit parameters according to eq.~(\ref{eq:SRatPol}) and eq.~(\ref{dagLi6}) for some relevant radiative capture reactions. $S_0$ is given in \si{\mega\electronvolt \milli\barn}, $a_k, q_k$ in $\si{(\mega\electronvolt)^{-k}}$.
	}
\end{table}

The only exception is the reaction $\pmb{d + \He{4} \to \Li{6} + \gamma}$, where a resonance appears. The $S$-factor here is given by
\begin{equation}\label{dagLi6}
	S(E) = S_0 \left( a_2 \cdot E^2 + \mathrm{BW}(E ; 1, \Gamma, M) \right) ,
\end{equation}
with the coefficients listed in Tabs.~\ref{tab:Par_RadCap} and~\ref{tab:Par_BW}.

\subsection{Charged particle reactions}

\begin{figure}
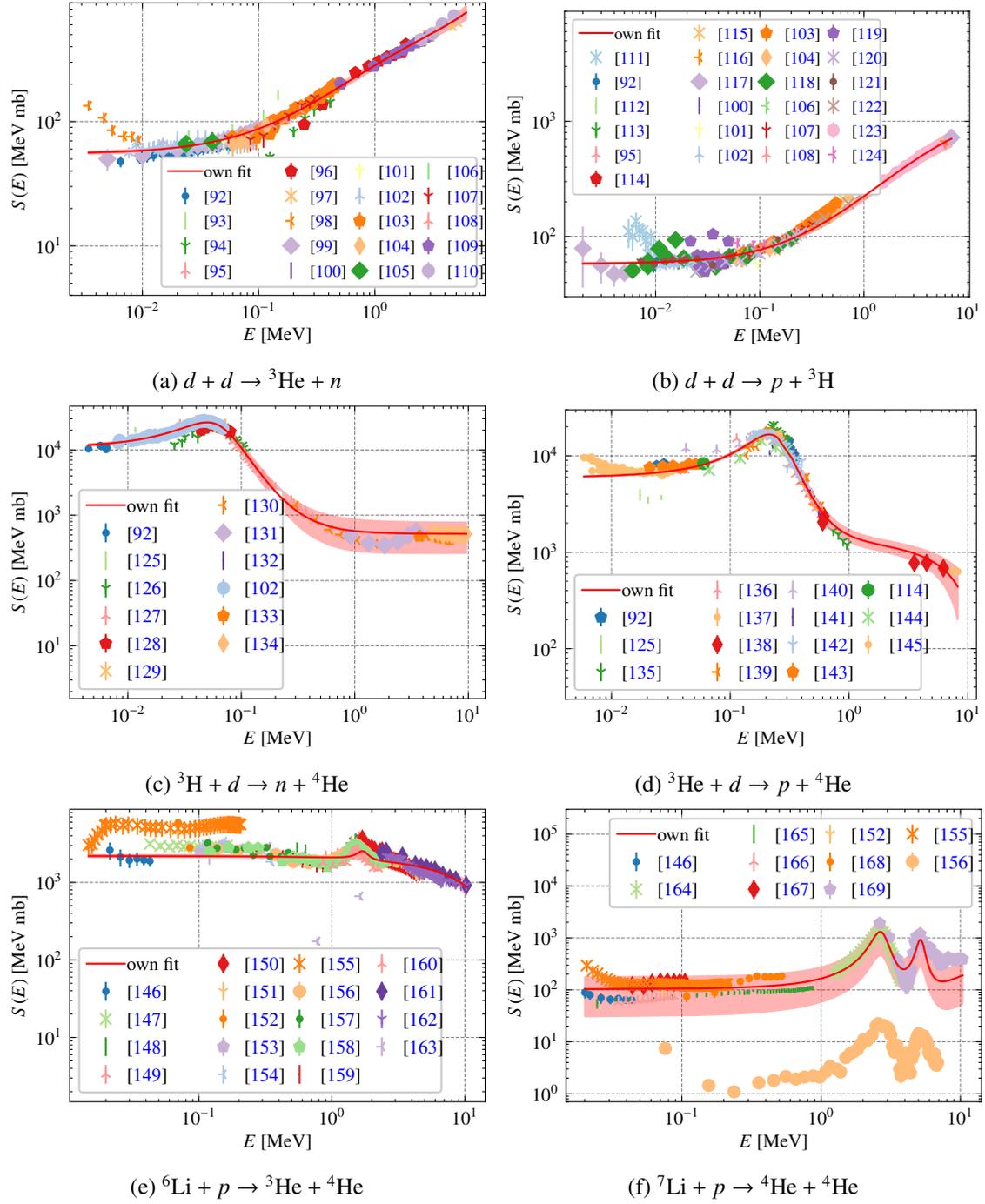

	\centering
	\begin{subfigure}{.49\textwidth}
		\resizebox{\textwidth}{!}{\input{figs/Params/ddnHe3_error.pgf}}
		\caption{$d + d \to \He{3} + n$}
	\end{subfigure}
	\begin{subfigure}{.49\textwidth}
		\resizebox{\textwidth}{!}{\input{figs/Params/ddpH3_error.pgf}}
		\caption{$d + d \to p + \triton$}
	\end{subfigure}
	\begin{subfigure}{.49\textwidth}
		\resizebox{\textwidth}{!}{\input{figs/Params/H3dnHe4_error.pgf}}
		\caption{$\triton + d \to n + \He{4}$}
	\end{subfigure}
	\begin{subfigure}{.49\textwidth}
		\resizebox{\textwidth}{!}{\input{figs/Params/He3dpHe4_error.pgf}}
		\caption{$\He{3} + d \to p + \He{4}$}
	\end{subfigure}
	\begin{subfigure}{.49\textwidth}
		\resizebox{\textwidth}{!}{\input{figs/Params/Li6paHe3_error.pgf}}
		\caption{$\Li{6} + p \to \He{3} + \He{4}$}
	\end{subfigure}
	\begin{subfigure}{.49\textwidth}
		\resizebox{\textwidth}{!}{\input{figs/Params/Li7paHe4_error.pgf}}
		\caption{$\Li{7} + p \to \He{4} + \He{4}$}
	\end{subfigure}
	
	\caption{Fits to experimental data composed by EXFOR \cite{Exfor_web} with error bands for some relevant charged particle reactions.}
	\label{fig:charged_par}
\end{figure}

\begin{figure}
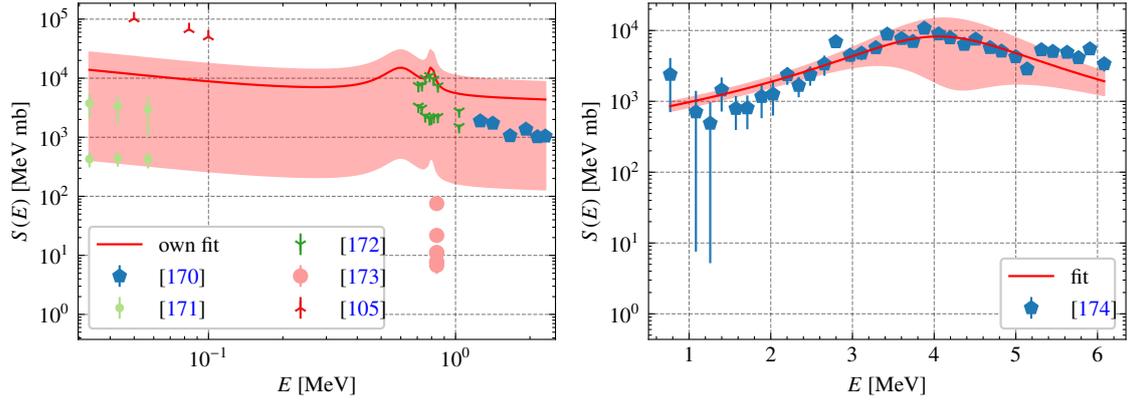

	\centering
	\begin{subfigure}{.49\textwidth}
		\resizebox{\textwidth}{!}{\input{figs/Params/Li7dnBe8_error.pgf}}
		\caption{$\Li{7} + d \to n + \He{4} + \He{4}$ ; error estimated to $\pm 97$.}
	\end{subfigure}
	\begin{subfigure}{.49\textwidth}
		\resizebox{\textwidth}{!}{\input{figs/Params/Be7dpBe8_error.pgf}}
		\caption{$\Be{7} + d \to p + \He{4} + \He{4}$}
	\end{subfigure}
	
	\caption{Fits to experimental data composed by EXFOR \cite{Exfor_web} with error bands for some relevant charged particle reactions with three particles in the final state.}
	\label{fig:charged_par2}
\end{figure}

As in Sect.~\ref{as:RadCap}, we fit an $S$-factor like Eq.~\eqref{eq:SRatPol}. The parameters are listed in Tab.~\ref{tab:Par_RadCap} for the charged particle reactions treated here. The $S$-factor fits are displayed with the data composed by \cite{Exfor_web} in Figs.~\ref{fig:charged_par} and~\ref{fig:charged_par2}.

\begin{table}[htb]
	\centering
	\sisetup{table-alignment=center, round-mode=places, round-precision=3, separate-uncertainty= true}
	\resizebox{\textwidth}{!}{%
	\begin{tabular}{@{}l%
			@{\extracolsep{1em}}S[table-format = 2.3e1,table-figures-uncertainty=1]%
			@{\extracolsep{0.5em}}S[table-format = 2.3,table-figures-uncertainty=1]%
			@{\extracolsep{0.5em}}S[table-format = 2.3,table-figures-uncertainty=1]%
			@{\extracolsep{0.5em}}S[table-format = 2.3,table-figures-uncertainty=1]%
			@{\extracolsep{0.5em}}S[table-format = 3.3,table-figures-uncertainty=1]%
			@{\extracolsep{0.5em}}S[table-format = 3.3,table-figures-uncertainty=1]%
			@{}}
		\toprule
	
		{Reaction}  & {$S_0$} & {$a_1$} & {$a_2$} & {$a_3$} & {$q_1$} & {$q_2$} \\
		\midrule
		
		{${d + d \to \He{3} + n}$} & 5.517 \pm 0.131 e1 & 6.701 \pm 0.377 & 0 & 0 & 0.531 \pm 0.044 & -0.032 \pm 0.005 \\  
		
		{${d + d \to p + \triton}$} & 5.786 \pm 0.088 e1 & 3.443 \pm 0.321 & 0 & 0 & 0.149 \pm 0.034 & 0 \\

		{$\triton + d \to n + \He{4}$} &  1.069 \pm 0.031 e4 & -0.994 \pm 1.342 & 11.819 \pm 5.744 & 0 & -24.260 \pm 0.444 & 243.991 \pm 8.442 \\

		{$\He{3} + d \to p + \He{4}$}  & 5.937 \pm 0.052 e3 & -2.049 \pm 0.175 & 3.835 \pm 0.296 & -0.305 \pm 0.063 & -6.788 \pm 0.053 & 15.604 \pm 0.223 \\

		{${\Li{6} + p \to \He{3} + \He{4}}$} & 2.162 \pm 0.076 e3   & -0.058 \pm 0.008 & 0 & 0 & 0 & 0\\
	
		{$\Li{7} + p \to \He{4} + \He{4}$} & -4.246 \pm 2.128 e1 & -0.522 \pm 0.222 & 0 & 0 & 0 & 0 \\
		
		{$\Li{7} + d \to n + \He{4} + \He{4}$} & 2.968 e3 & 8.27894999781032 & -0.3081440001078 & 0 & 54.611 & 0 \\
		
		{$\Be{7} + d \to p + \He{4} + \He{4}$} & 5.817 \pm 0.843 e2 & 0 & 0 & 0 & -0.458 \pm 0.010 & 0.057 \pm 0.002 \\
		
		\bottomrule
	\end{tabular}}
	\caption{  \label{tab:Par_Charged}
		Fit parameters according to eqs.\,(\ref{eq:SRatPol}), (\ref{Li67p}) and (\ref{Li7d}) for charged particle reactions. $S_0$ is given in \si{\mega\electronvolt \milli\barn}, $a_k, q_k$ in $(\si{\mega\electronvolt})^{-k}$.
	}
\end{table}

For the reactions $\pmb{\Li{6} + p \to \He{3} + \He{4}}$ and $\pmb{\Li{7} + p \to \He{4} + \He{4}}$ which have one (two) resonance(s), the $S$-factor is given by
\begin{equation}\label{Li67p}
	S(E) =  S_0 \left(1 + a_1 E + a_2 E^2 + a_3 E^3 + \mathrm{BW}(E ; k_1, \Gamma_1, M_1) + \mathrm{BW}(E ; k_2, \Gamma_2, M_2)  \right). 
\end{equation}

For the reaction $\pmb{\Li{7} + d \to n + \He{4} + \He{4}}$, we use the parameterization from~\cite{Meissner:2022dhg} which has the form
\begin{equation}\label{Li7d}
		S(E) = S_0 \frac{1 + a_1 E + a_2 E^2}{1 + q_1 E} + \mathrm{bw}(E ; k_1, \kappa_1, M_1) + \mathrm{bw}(E ; k_2, \kappa_2, M_2),
\end{equation}
and estimate the error to be $\pm 97\%$ (so that almost all data points are included).

\subsection{Neutron-induced reactions}

\begin{figure}
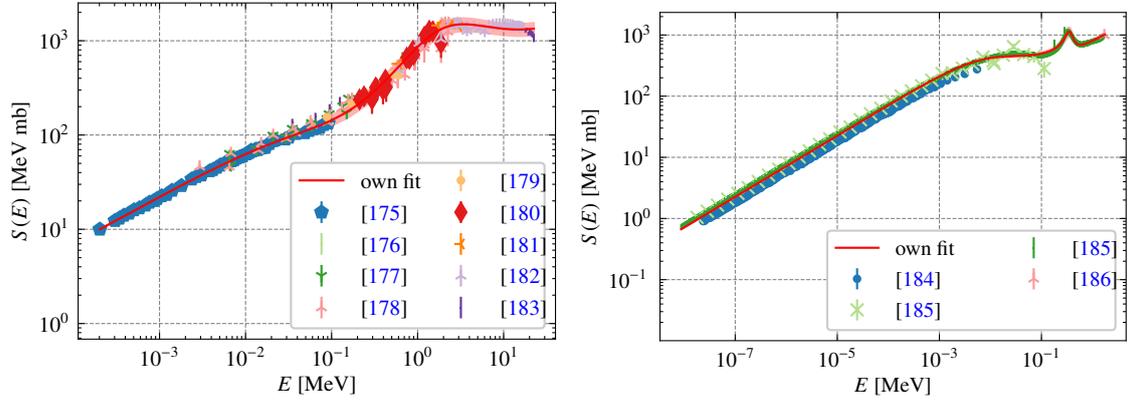
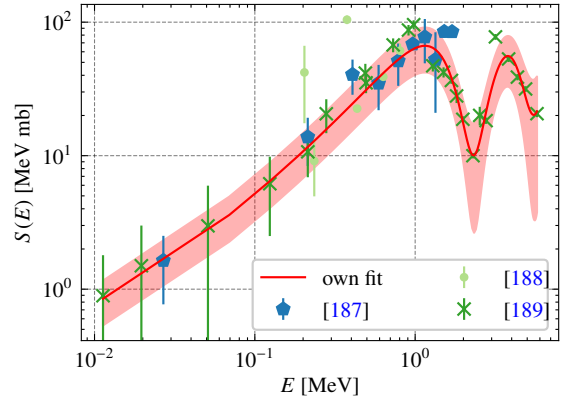

	\centering
	\begin{subfigure}{.49\textwidth}
		\resizebox{\textwidth}{!}{\input{figs/Params/He3npH3_error.pgf}}
		\caption{$\He{3} + n \to p + \triton$}
	\end{subfigure}
	\begin{subfigure}{.49\textwidth}
		\resizebox{\textwidth}{!}{\input{figs/Params/Be7npLi7_error.pgf}}
		\caption{$\Be{7} + n \to p + \Li{7}$}
	\end{subfigure}
	\begin{subfigure}{.49\textwidth}
		\resizebox{\textwidth}{!}{\input{figs/Params/Be7naHe4_error.pgf}}
		\caption{$\Be{7} + n \to \He{4} + \He{4}$ ; error estimated using only $\Delta S_0$ and $\Delta a_1$.}
	\end{subfigure}
	\caption{Fits to experimental data composed by EXFOR \cite{Exfor_web} with error bands for some relevant neutron-induced reactions.}
	\label{fig:neutron_par}
\end{figure}

The parameters found for three neutron-induced reactions parameterized here are displayed in Tab.~\ref{tab:Par_Neutron} and compared to data composed by \cite{Exfor_web} in Fig.~\ref{fig:neutron_par}.

\begin{table}[htb]
	\centering
	\sisetup{table-alignment=center, round-mode=places, round-precision=3, separate-uncertainty=true}
 	\resizebox{\textwidth}{!}{%
 	\begin{tabular}{@{}l%
			@{\extracolsep{1em}}S[table-format = 2.3e1,table-figures-uncertainty=1]%
			@{\extracolsep{0.5em}}S[table-format = 2.3,table-figures-uncertainty=1]%
			@{\extracolsep{0.5em}}S[table-format = 2.3,table-figures-uncertainty=1]%
			@{\extracolsep{0.5em}}S[table-format = 1.3,table-figures-uncertainty=1]%
			@{\extracolsep{0.5em}}S[table-format = 2.3,table-figures-uncertainty=1]%
			@{\extracolsep{0.5em}}S[table-format = 2.3,table-figures-uncertainty=1]%
			@{\extracolsep{0.5em}}S[table-format = 1.3,table-figures-uncertainty=1]%
			@{}}
		\toprule
		{Reaction}  & {$\mathcal{R}_0$} & {$a_1$} & {$a_2$} & {$a_3$} & {$q_1$} & {$q_2$} & {$q_3$} \\
		\midrule
	
		$\He{3} + n \to p + \triton$  & 7.053 \pm 0.034 e2 & 8.359 \pm 1.763 & 22.042 \pm 2.556 & 1.571 \pm 0.273 & 22.192 \pm 2.708 & -2.601 \pm 2.445 & 6.462 \pm 0.616 \\

		{$\Be{7} + n \to p + \Li{7}$}  & 6.962 \pm 0.035 e3 & 9.375 \pm 0.488 & 0 & 0 & 89.309 \pm 3.542 & 0 & 0 \\

		{$\Be{7} + n \to \He{4} + \!\!\He{4}$}  & -2.765 \pm 1.01 e2 & -0.160 \pm 0.010 & 0 & 0 & 0 & 0 & 0 \\
		\bottomrule
	\end{tabular}}
	\caption{  \label{tab:Par_Neutron}
		Fit parameters according to Eq.~(\ref{eq:R_NC}) for neutron-induced reactions. $\mathcal{R}_0$ is given in \si{(\mega\electronvolt)^{1/2} \milli\barn}, $a_k, q_k$ in $(\si{\mega\electronvolt})^{-k}$.
	}
\end{table}

The parameterization for the reactions $\pmb{\Be{7} + n \to \Li{7} + p}$ (one resonance) and $\pmb{\Be{7} + n \to \He{4} + \He{4}}$ (two resonances) is of the form
\begin{equation}\label{Be7naa}
\mathcal{R}(E) = \mathcal{R}_0 \left( \frac{1 + a_1 E + a_2 E^2}{1 + q_1 E}  + \mathrm{BW}(E;k_1, \Gamma_1, M_1) + \mathrm{BW}(E;k_2, \Gamma_2, M_2) \right).
\end{equation}

\begin{table}
	
	\centering
	\sisetup{table-alignment=center, round-mode=places, round-precision=3, separate-uncertainty=true}
	\resizebox{\textwidth}{!}{%
		\begin{tabular}{@{}l%
				@{\extracolsep{1em}}S[table-format = 4.3e2,table-figures-uncertainty=1]%
				@{\extracolsep{0.5em}}S[table-format = 2.3,table-figures-uncertainty=1]%
				@{\extracolsep{0.5em}}S[table-format = 1.3,table-figures-uncertainty=1]%
				@{\extracolsep{0.5em}}S[table-format = 4.3,table-figures-uncertainty=1]%
				@{\extracolsep{0.5em}}S[table-format = 4.3,table-figures-uncertainty=1]%
				@{\extracolsep{0.5em}}S[table-format = 1.3,table-figures-uncertainty=1]%
				@{}}
			\toprule
			{Reaction} & {$k_1$} & {$\Gamma_1$} & {$M_1$} & {$k_2$} & {$\Gamma_2$} & {$M_2$} \\ \midrule
			{$d + \He{4} \to \Li{6} + \gamma$} & 1 & .028 \pm 0.007 & .711 \pm 0.002 &  &  &  \\ 
			
			{$\Li{6} + p \to \He{3} + \He{4}$} & 0.130 \pm 0.088 & 0.397 \pm 0.135 & 1.708 \pm 0.042 &  &  &  \\
			{$\Li{7} + p \to \He{4} + \He{4}$} & -398.050 \pm 189.573 & 0.866 \pm 0.098 & 5.171 \pm 0.032 & -159.822 \pm 64.838 & 0.872 \pm 0.060 & 2.657 \pm 0.017  \\ 
			
			{$\Be{7} + n \to p + \Li{7}$} & 4.430 \pm 0.923 e-4 & 0.162 \pm 0.020 & 0.341 \pm 0.006 &  &  & \\
			
			 {$\Be{7} + n \to \He{4} + \He{4}$}  & -92.818 \pm 35.910 & 3.644 \pm 0.812 & 3.450 \pm 0.118 & -5.969 \pm 2.532 & 3.253 \pm 1.544 & 0.872 \pm 0.218 \\ 
			\midrule\midrule
			 & {$k_1$} & {$\kappa_1$} & {$M_1$} & {$k_2$} & {$\kappa_2$} & {$M_2$}  \\ \midrule
			{$\Li{7} + d \to n + \He{4} + \He{4}$} & 9820.6 & 82.3871  & 0.6 & 8990.99 & 1963.84 & 0.8   \\ 
			\bottomrule
		\end{tabular}
	}
	\caption{
		\label{tab:Par_BW}
		Fit parameters for resonances parameterized as Breit-Wigner functions (Eq.~\eqref{eq:BWfunc} and Eq.~(\ref{eq:BWfunc_nr}). $\Gamma_k, M_k$ are given in \si{\mega\electronvolt}, $\kappa_k$ in \si{(\mega\electronvolt)^{-2}}. For the reaction $\Li{7} + d \to n + \He{4} + \He{4}$ $k$ is given in $\si{\mega\electronvolt \milli\barn}$, otherwise it is dimensionless.
	}
\end{table}

\newpage
\bibliographystyle{JHEP} 
\bibliography{refs}

\end{document}